\documentclass[
aps,prd,
nofootinbib,
superscriptaddress,
showpacs,
tightenlines,
]{revtex4}
\usepackage{amsmath}
\usepackage{amssymb}
\usepackage{bm}
\usepackage{color,graphicx}
\usepackage{epsfig}

\begin{document}

\title{Interpretation of the Flavor Dependence of Nucleon Form Factors in a Generalized Parton Distribution Model}

\author{J. Osvaldo Gonzalez-Hernandez} 
\email{jog4m@virginia.edu}
\affiliation{Department of Physics, University of Virginia, Charlottesville, VA 22904, USA.}

\author{Simonetta Liuti} 
\email{sl4y@virginia.edu}
\affiliation{Department of Physics, University of Virginia, Charlottesville, VA 22904, USA}
\affiliation{Laboratori Nazionali di Frascati, INFN, Frascati, Italy}

\author{Gary R.~Goldstein} 
\email{gary.goldstein@tufts.edu}
\affiliation{Department of Physics and Astronomy, Tufts University, Medford, MA 02155 USA.}

\author{Kunal Kathuria} 
\email{kk7t@virginia.edu}
\affiliation{Department of Physics, University of Virginia, Charlottesville, VA 22904, USA.}


\begin{abstract}
We give an interpretation of  the $u$ and $d$ quarks contributions to the nucleon electromagnetic form factors  for values of the four-momentum transfer in the multi-GeV region where flavor separated data have been recently  made available.  The data show, in particular, a suppression of $d$ quarks with respect to $u$ quarks at large momentum transfer.
This trend can be explained using a reggeized  diquark model calculation of generalized parton distributions, thus providing a correlation between momentum and coordinate spaces, both of which are necessary in order to interpret the partonic substructure of the form factors.  We extend our discussion to the second moments of generalized parton distributions which are believed to contribute to partonic angular momentum.
\end{abstract}

\maketitle

\baselineskip 3.0ex
\section{Introduction}
A recent experimental analysis displays a flavor separation of the nucleon elastic electromagnetic form factors \cite{Cates}. The study in \cite{Cates} brings further and completes various previous analyses nicely summarized in Ref.\cite{Qattan} (see in particular \cite{Craw,Rio}). 
The data show a suppression of the contribution of $d$ quarks  with respect to $u$ quarks at large momentum transfer. The suppression is observed in both the Dirac, $F_1^{q=u,d}$,  and Pauli, $F_2^{q=u,d}$, form factors. 
In particular, for  four-momentum transfer squared, $1.5 \lesssim -t \lesssim 4$ GeV$^2$, the $d$ quarks form factors fall as $1/(-t)^2$, while the $u$ quarks' fall as $1/(-t)$. 

A quantitative study connecting the fall-off  of the form factor with $-t$ and  the radii of partonic configurations in the nucleon was performed in Ref.\cite{Roberts}  
where it was suggested that a steeper slope of the $d$ vs. $u$ contributions can be attributed to the quark-diquark structure of the proton. In a nutshell, according to \cite{Roberts} a struck $d$ quark leaves behind a larger mass  axial-vector diquark  and is therefore positioned  further away, in average, from the quark-diquark system's center of mass as compared to a struck $u$ quark leaving behind a smaller mass scalar diquark. 

 It is well known that a  description of the  relativistic 3D structure of hadrons in coordinate space applies within a well defined range of validity. In fact, the most recent  precise estimate given in  \cite{Roberts} shows that nucleon substructure up to $ \approx 81  \%$ of the nucleon volume can be accounted for at $\sqrt{-t}  \lesssim1$ GeV. 
Away from the non-relativistic limit,  the motion of the center of mass cannot be separated straightforwardly from the relative motion (see the review in \cite{Miller_annrev}), and ``recoil" corrections become important.
 In Ref.\cite{Soper77} it was shown, however, that by adopting a light-front framework, where partons are specified by their ``+"  momentum component (where we define, $k^\pm=(k_o\pm k_3)/\sqrt{2}$), and by their transverse coordinate, ${\bf x}$, one obtains a relation between the transverse parton density in  coordinate space and the form factors  that is analogous to the non relativistic one. The reason behind this simplification is that in the parton picture, or on the light-front, the subgroup of the Poincar\'{e} group that leaves the $x^+=0$ surface constant is isomorphic to the Galileian group in the transverse plane. Based on this observation, Soper introduced the impact parameter dependent parton distribution functions  \cite{Soper77}, $q(x,{\bf b})$ where $x=k^+/P^+$ is the parton's light cone momentum fraction, and ${\bf b}$ (related to ${\bf x}$ \cite{Soper77}), defines its transverse distance  from the center of ``+"  momentum. 
  
It was thanks to this step that, soon after the introduction of Generalized Parton Distributions (GPDs)  \cite{GPD1,GPD2,GPD3},  Burkardt \cite{Burkardt_radius} suggested a  connection between transverse coordinate dependent parton densities and 
observables from Deeply Virtual Compton Scattering (DVCS) and related experiments. 
In Ref.\cite{Burkardt_radius} the GPDs were shown to be the 2D Fourier transforms of the impact parameter dependent  parton distribution functions, $q(x,{\bf b})$.  Complementary information was subsequently obtained in Ref.\cite{Miller_1} where the link between the 3D coordinate space density $\rho(x^-,{\bf b})$ to the 2D,  transverse density $\rho({\bf b})$, was defined, with $\rho({\bf b}) = \int dx \, q(x,{\bf b})$.
 
%
\begin{figure}
\includegraphics[width=8.cm]{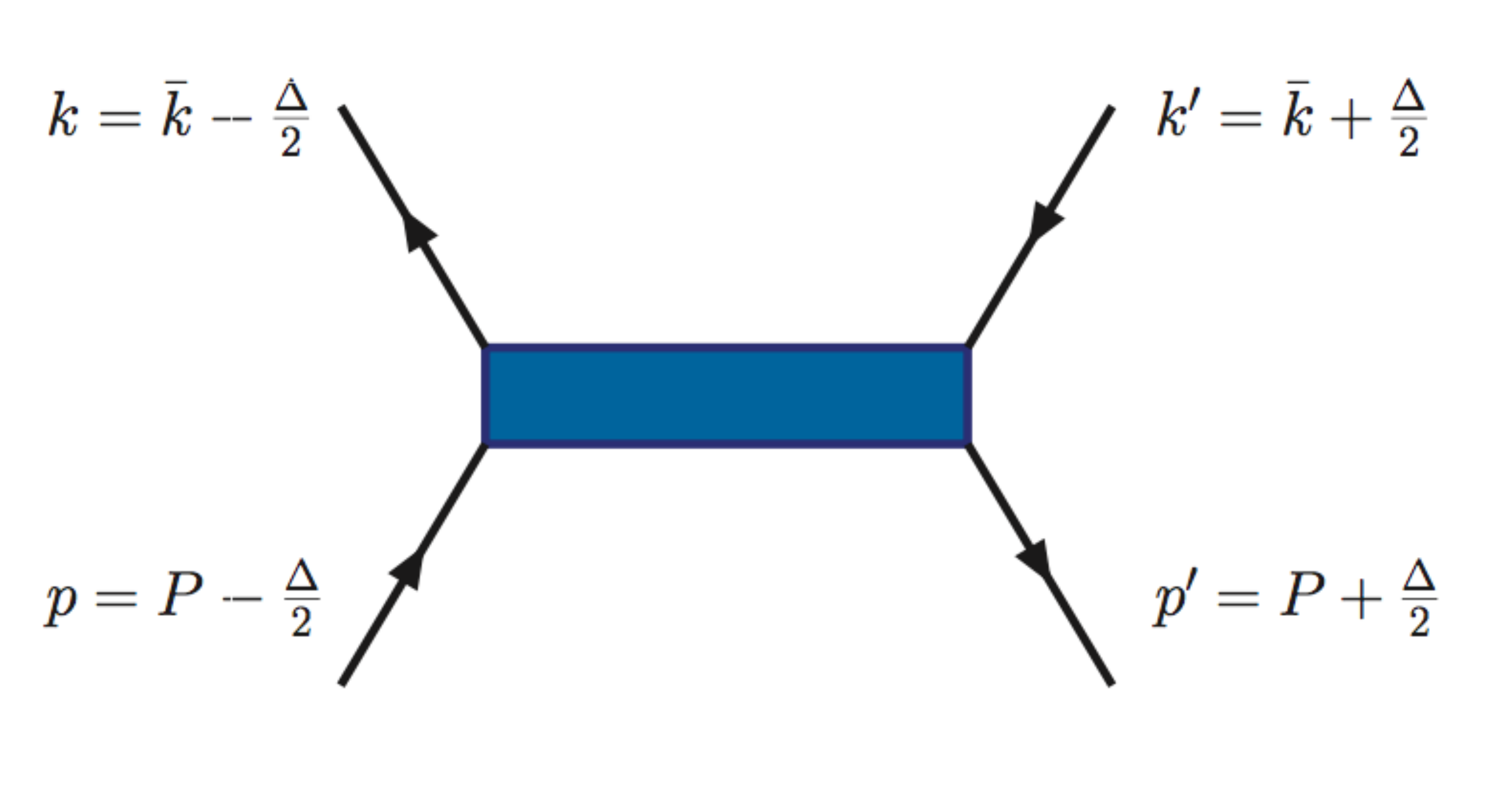}
\caption{GPDs kinematics}
\label{fig1}
\end{figure}

In this paper we study the flavor dependence of the form factors by using the connection with GPDs provided by the following sum rules \cite{GPD2} and Fourier transforms \cite{Burkardt_radius} respectively,
 \begin{eqnarray}
\label{formf}
F_1^q(t) & = & \int_0^1 dx H_q(x,\xi,t), \;\;\;\;  F_2^q(t)  =  \int_0^1 dx E_q(x,\xi,t) \\
\label{FT}
H_q(x,0,t) & = & \int d^2{\bf b}  \, \rho_q(x,{\bf b}) \, e^{i {\bf b}\cdot {\bf \Delta}_\perp},   \;\;\; E_q(x,0,t)  =  \int d^2{\bf b}  \, {\cal E}_q(x,{\bf b}) \, e^{i {\bf b}\cdot {\bf \Delta}_\perp}
\end{eqnarray}
where  $H_q$ and $E_q$, $q=u,d$, are the GPDs,%
\footnote{Strictly the valence quarks GPDs, see {\it e.g.} Ref.\cite{hybrid_even}}
 which depend on the partons' momenum fractions $x=\bar{k}^+/P^+$, the skewness, $\xi=\Delta^+/2P^+$, and the invariant $t =\Delta^2$, $\Delta$ being the four-momentum transfer between the initial and final proton (see Fig.\ref{fig1} and Refs.\cite{Diehl_rev,BelRad} for a review).
Lorentz invariance implies that the first moment of GPDs in the parton's longitudinal momentum fraction, $x$, defining the form factors, Eq.(\ref{formf}), is $\xi$ independent. 

GPDs describe hybrid properties of the deeply virtual structure of nucleons \cite{Diehl_rev}. 
Eq.(\ref{formf}), in particular, establishes a connection between the deep inelastic structure and the form factor of the proton. This information is contained in the integrand, the GPD, which, it is important to stress, is itself an observable that can be extracted from a different, independent, set of measurements.%
\footnote{More precisely GPDs are contained in the directly measurable Compton factors \cite{Diehl_rev,BelRad}.} 

By Fourier transforming $H_q(x,0,t)$ in the variable ${\bf \Delta}_\perp$ ($t=-{\bf \Delta}_\perp^2$), as in Eq.(\ref{FT}), one obtains a single-particle density, or a diagonal object in impact parameter space, $\rho_q(x,{\bf b})$ (similarly, for $E_q$) \cite{Burkardt_radius}. 
We will therefore focus from now on,  on the zero skewness components of the GPDs although considering models that satisfy the polynomiality property by construction.
The formal backbone to this picture -- the dominance of the handbag diagram -- is provided by well established factorization theorems \cite{GPD2,ColFraStr}. 
QCD-based models giving an interpretation of the parton correlators for both exclusive and inclusive processes in terms of their dominant degrees of freedom  ({\it e.g.} quark or diquark correlations) are,  however, debatable. 
Reviews on both the history and the more recent important developments on this longstanding issue, in the elastic scattering sector, can be found  {\it e.g.} in Refs.\cite{Roberts,Arrington,Gao}. 

Our analysis is based on the ``flexible" parametrization introduced in Ref.\cite{hybrid_even} where using a reggeized quark diquark model we provided a quantitative fit  the proton's and neutron's electroweak form factors. The large number of parameters that is necessary to fit GPDs was handled by using a recursive procedure. 

An essential component of our approach was using the form factor data in order to constrain the $t$ dependence of the GPDs, as shown by Eq.(\ref{formf}) for the electromagnetic sector. 
We now ask the question of what components  of  the nucleon's partonic substructure that characterize our model of GPDs, allow us to reproduce the form factors to high accuracy. 

In order to expound this question we proceed stepwise. Our paper is organized as follows: in Section \ref{sec:2} we summarize our GPD model, focusing on those aspects that impact directly the form factor flavor structure; in Section \ref{sec:3} we give an interpretation of the form factors flavor dependence, and we present our prediction for the GPDs second moments, $J_u$, and $J_d$; in Section \ref{sec:4} we draw our conclusions.

\section{Reggeized Diquark Model} 
\label{sec:2}
\begin{figure}
\includegraphics[width=9.cm]{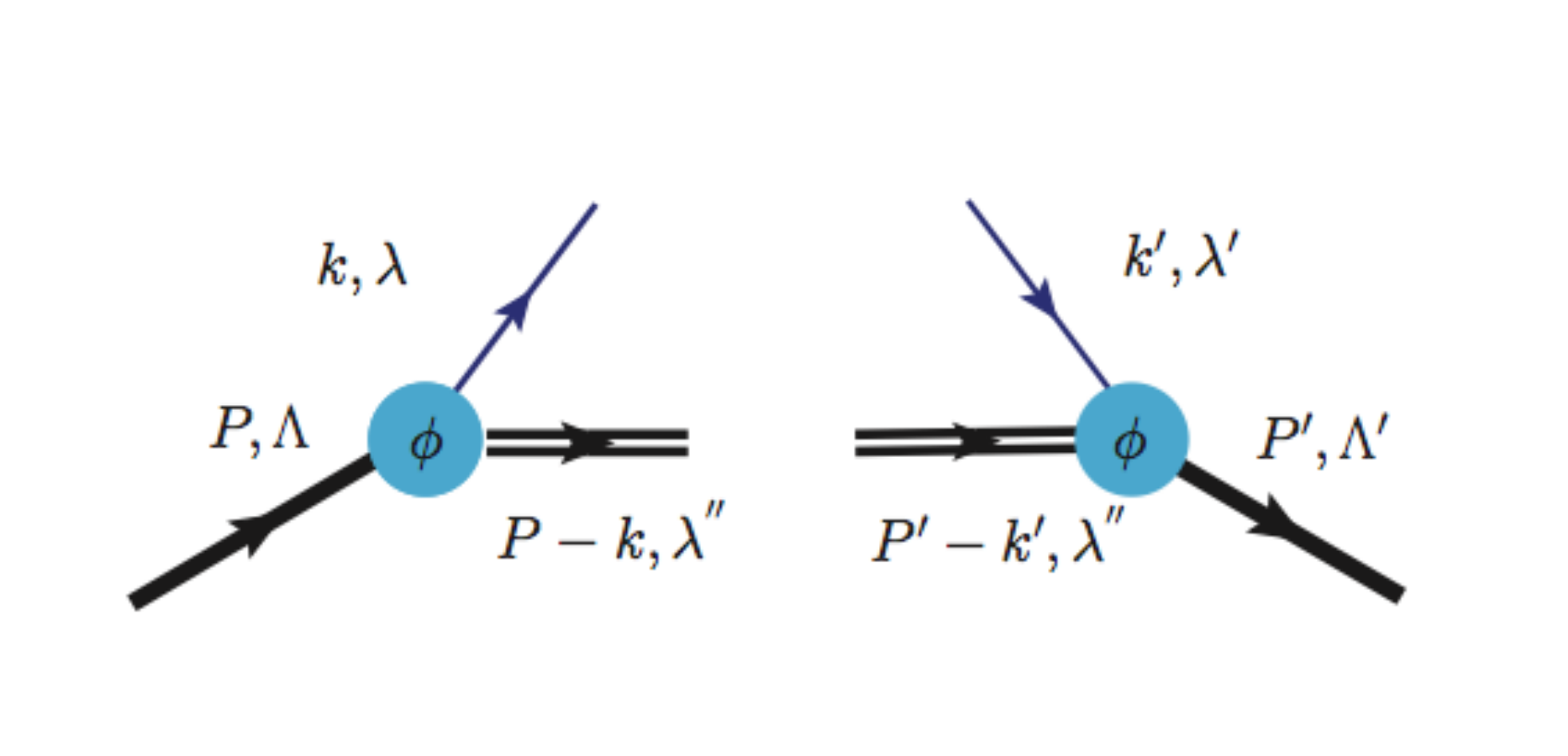}
\caption{Vertex structures defining the spectator model tree level diagrams.  
}
\label{fig2}
\end{figure}

\begin{figure}
\large{\bf (a)}
\includegraphics[width=5.5cm,angle=0]{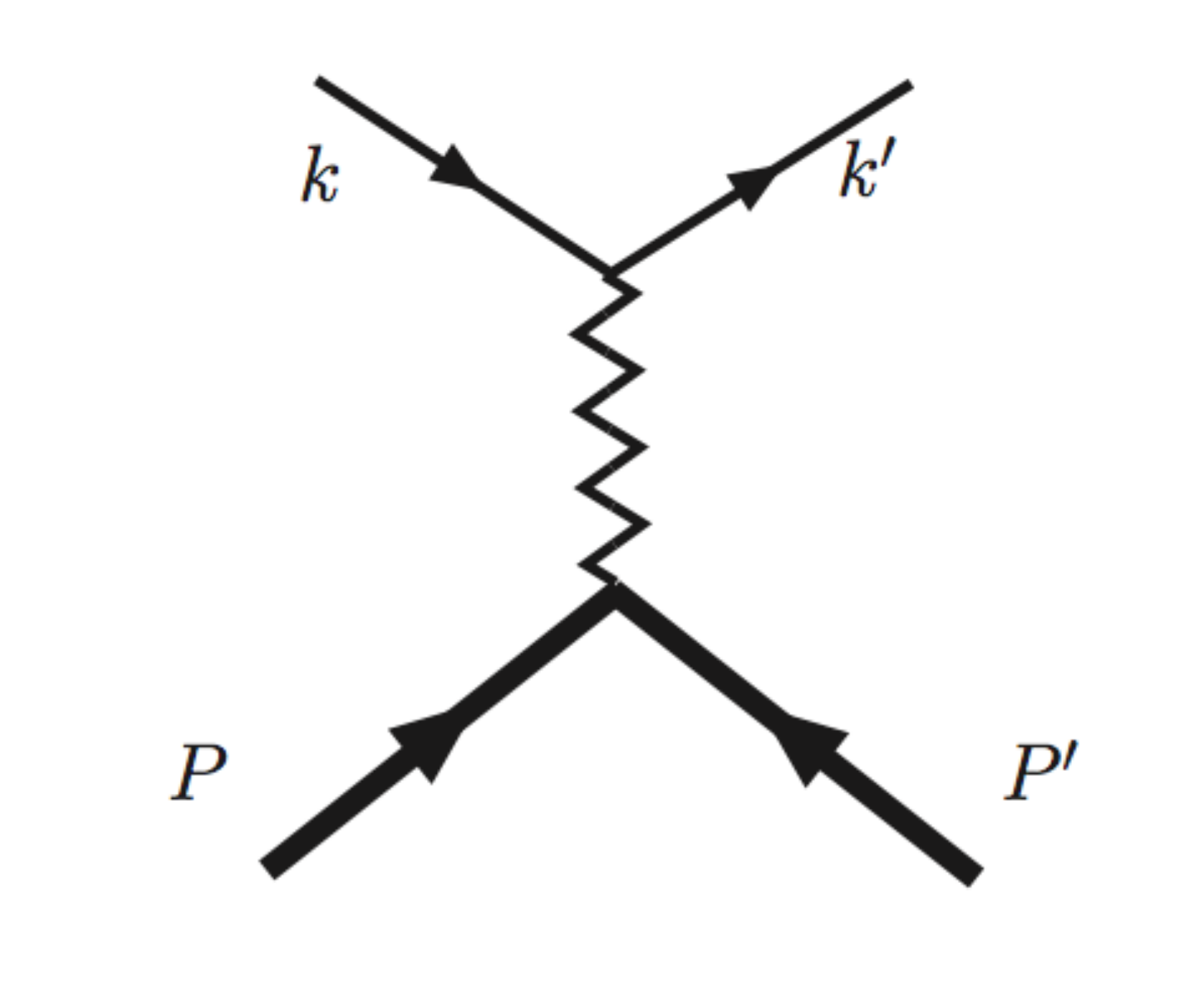}
\large{\bf (b)}
\includegraphics[width=7.5cm,angle=0]{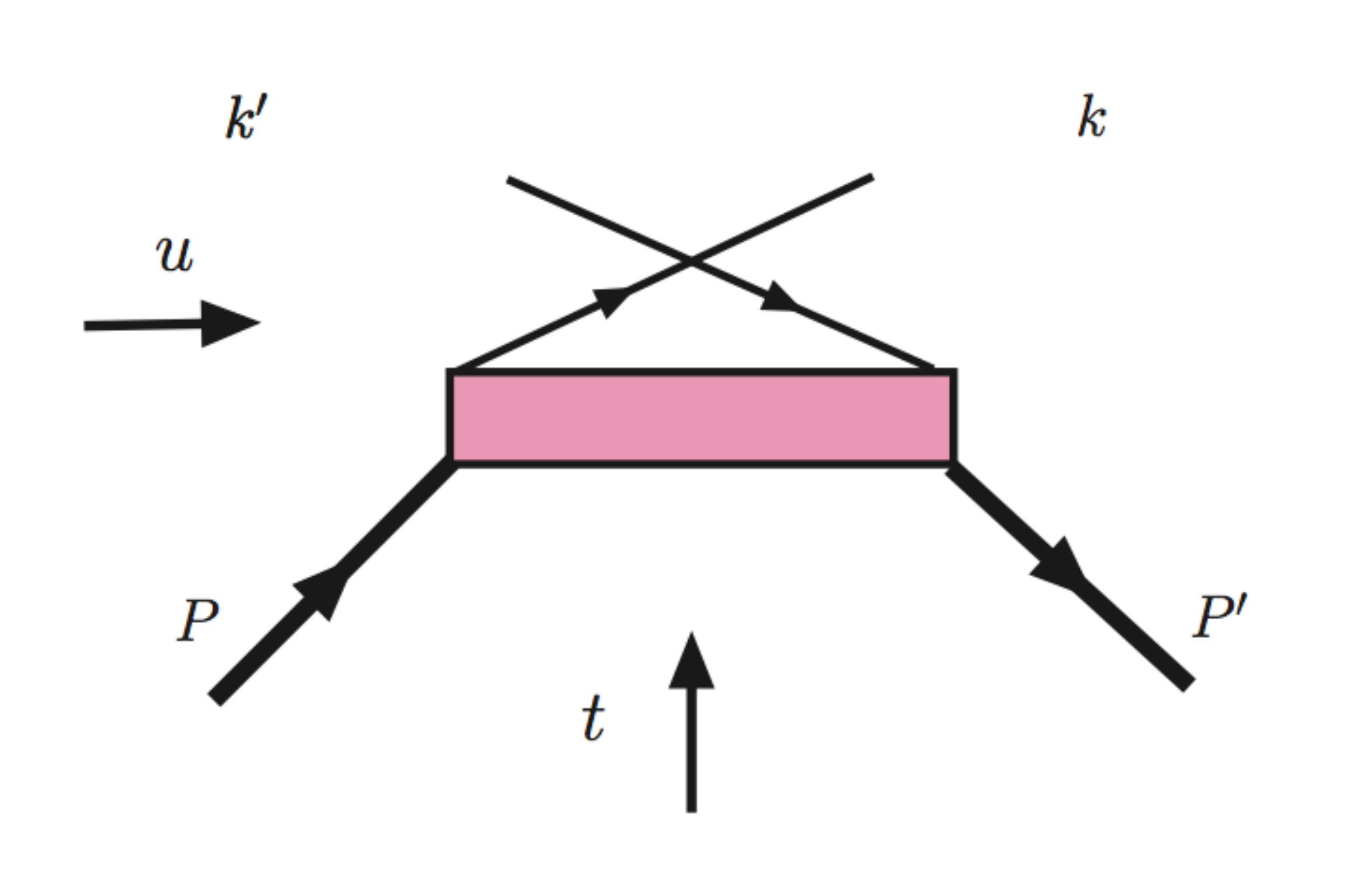}
\normalsize
\caption{(a) $t$-channel Reggeon exchange diagram; (b) $u$-channel diquark exchange. The box has mass $M_X$, with spectral distribution $\rho(M_X^2)$ as described in the text.}
\label{fig3}
\end{figure}

In Refs.\cite{hybrid_even,AHLT1,AHLT2} we developed a quark diquark  model with the aim of interpreting DVCS data. 
The basic structures of the model are the helicity quark-proton scattering amplitudes at leading order with proton-quark-diquark vertices (Fig.\ref{fig2}).
The dominant components are quark-diquark correlations where the diquark system has both a finite radius and an invariant mass, $M_X$, that is let vary according to a spectral distribution, differently from most models where the recoiling system's mass is kept fixed \cite{BCR,CloetMiller}. The variable mass diquark systems exhibit different structure as one goes from low to high mass values: at low mass values one has a simple two quarks system composed with spin $J=0^+,1^+$, whereas at large mass values more complex correlations ensue which are regulated by the Regge behavior of the quark-proton amplitude, $\propto \hat{u}^{\alpha(t)} = (M_X^2)^{\alpha(t)}$, Fig.\ref{fig3}.  
This behavior, also known as reggeization (see Ref.\cite{Forshaw}, Ch. 3 and references therein), is regulated by a spectral distribution, $\rho(M_X^2)$. 
We will show how upon integration over the mass, the spectral distribution yields on one side for small $x$  the desired $x^{-\alpha}$ behavior, and on the other for intermediate and large $x$, it is consistent with the diquark model. 

The need for introducing a Regge term while applying diquark models to GPDs  was realized in previous phenomenological studies  \cite{AHLT1,AHLT2}. It was noticed that while it is a known fact that the diquark model cannot produce a steep enough increase of the structure functions at low $x$, and this might be of minor importance in kinematical regions centered at relatively large $x$ where most data in the multi-GeV region are; it is, however, a necessary contribution to obtain the normalization of the structure functions correctly. This observation becomes important for GPDs where we require them to reproduce the form factor's behavior exactly through their normalization, or first moment as given in Eq.(\ref{formf}). The Regge term is therefore an essential ingredient in model building. The importance of the Regge term was realized recently also in Refs.\cite{Rad11,Szcz11}. There, however, the more singular behavior is introduced with a slightly different motive -- it is required in order to model GPDs from a single Double Distribution.      

In practical terms,  our reggeized diquark model depends on a number of parameters that we divide into  Regge and pure diquark contributions. 
The parameters were fixed by a fit applied recursively to PDFs from deep inelastic scattering data,  and to form factors and DVCS data from Jefferson Lab \cite{HallB}. The model was subsequently compared  to data on different observables (charge and transverse single spin asymetries), in a different kinematical regime from HERMES \cite{HERMES1,HERMES2}. We define our parametrization as ``flexible" in that, mostly owing to its recursive feature, the different components can be efficiently fitted 
 separately as new data come in.  

Summarizing, we consider Ref.\cite{hybrid_even} as the accomplishment of a first phase in which we constructed a reggeized diquark  model which satisfies fundamental requirements, such as polynomiality, positivity, crossing symmetries, hermiticity, and time reversal invariance. 
In the process, we studied the behavior of the various parameters  both for the forward limit and for the integral relations including form factors, and we reproduced a number of observables. 
Our main result is summarized in Table I of Ref.\cite{hybrid_even} where an optimal set of the parameters obtained from data available at the time of publication was  presented.  
While the parameters in the diquark part of our model are essentially masses that have precise definitions and boundaries for their values which have been addressed extensively in the literature (see for instance \cite{BCR}), the physical origin of the $t$ dependence stemming from the term displaying Regge behavior has not been sufficiently discussed.   
In what follows we, therefore, contribute a discussion of this term.

The fitting procedure of GPDs is quite complicated because of the presence of many different steps. In Fig.\ref{fig:flow} we present a flowchart that both summarizes and streamlines the various steps described so far.

\subsection{Description of parameters}
\begin{figure}
\includegraphics[width=16.0cm]{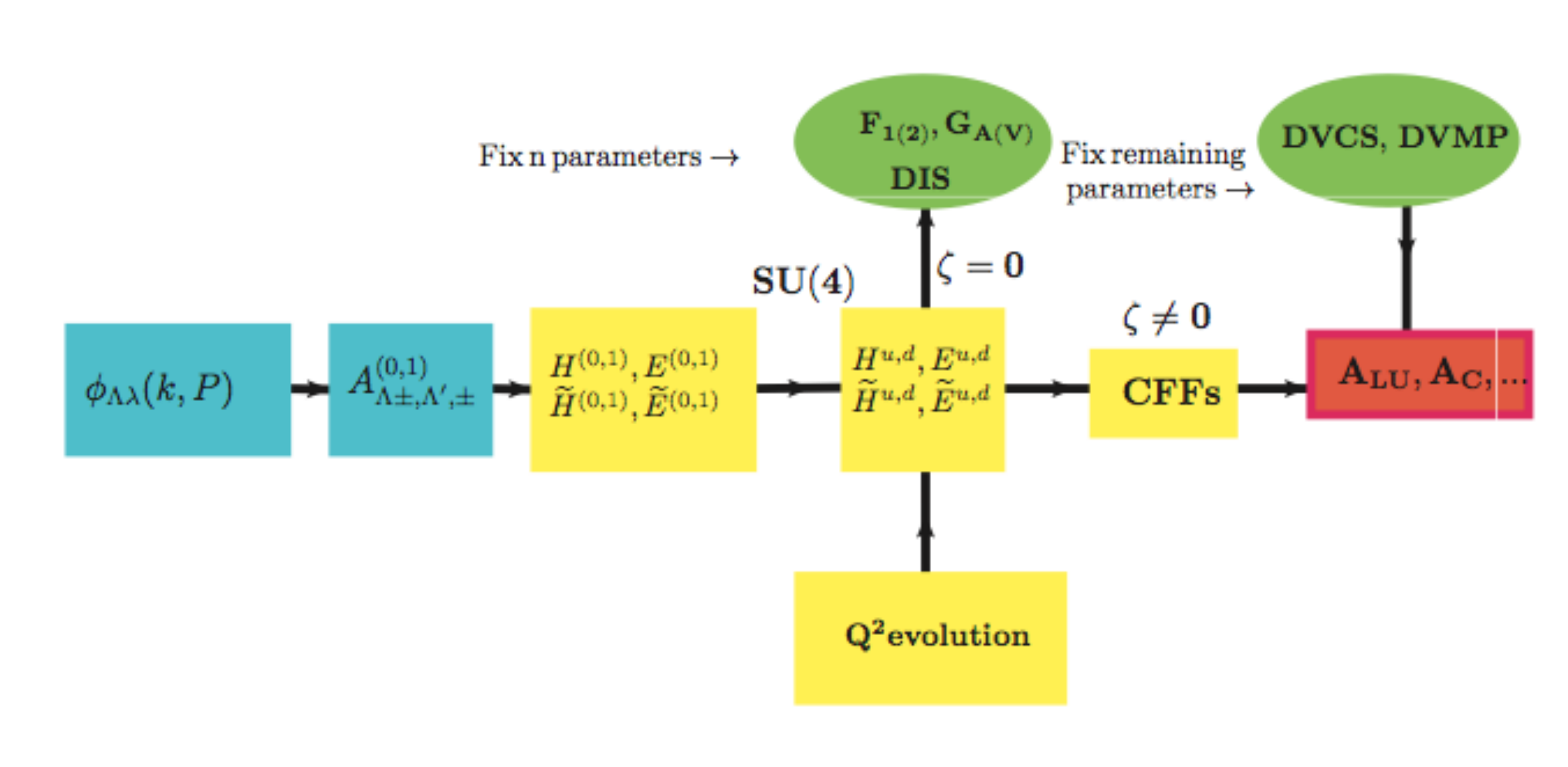}
\caption{(Color online) Flowchart for the GPD fitting procedure described in the text.}
\label{fig:flow}
\end{figure}

Our GPD model can be summarized in the following expression, 
\begin{equation}
F_q(X,\zeta,t)  = {\cal N}_q G_{M_X^q,m_q}^{M_\Lambda^q}(X,\zeta,t) \,  
R^{\alpha_q,\alpha_q^\prime}_{p_q}(X,\zeta,t) 
\label{fit_form}
\end{equation}
where $q=u,d$, $F_q \equiv H_q, E_q$; the functions $G_{M_X^q,m_q}^{M_\Lambda^q} \equiv H_{M_X^q,m_q}^{M_\Lambda^q}, E_{M_X^q,m_q}^{M_\Lambda^q}$, and  
$R^{\alpha_q,\alpha_q^\prime}_{p_q} \equiv R_{H(E), p_q}^{\alpha_q,\alpha_q^\prime}$, parametrize respectively, the quark-diquark and Regge contributions; $X$ and $\zeta$, the variables in the asymmetric system \cite{Diehl_rev}, are related to 
$x$ and $\xi$ by, $X=(x+\xi)/(1+\xi)$, $\xi=2\zeta/(2-\zeta)$, however, here we are interested in the $\zeta=\xi=0 \Rightarrow X=x$ limit. The diquark components read, 
\begin{subequations}
\label{G}
\begin{eqnarray}
\label{H}
H_{M_X^q,m_q}^{M_\Lambda^q}  & = &  \displaystyle\mathcal{N}_q   \int \frac{d^2k_\perp}{1-x}  \; \, 
\frac{%
\left[  \left(m_q+M x\right)  \left(m_q + M x \right) + {\bf k}_\perp\cdot \tilde{{\bf k}}_\perp\right]}
{[ {\cal M}_q^2(x)  - k_\perp^2/(1-x) ]^2 [{\cal M}_q^2(x)  - \tilde{ k}^2_\perp/(1-x)]^2 } \\
\label{E}
E_{M_X^q,m_q}^{M_\Lambda^q}  & = & \displaystyle \mathcal{N}_q \int  \frac{d^2 k_\perp}{1-x}  \; \, 
 \frac{-2M /{\Delta_\perp^2} %
 \left[  \left(m_q+M x \right) \displaystyle   {\bf \tilde{k}}_\perp \cdot {\bf \Delta}_\perp  - \left(m_q + M x \right) {\bf  k}_\perp \cdot {\bf \Delta}_\perp  \right] }
 {[ {\cal M}_q^2(x)  - k_\perp^2/(1-x) ]^2 [{\cal M}_q^2(x)  - \tilde{ k}^2_\perp/(1-x)]^2 } 
\end{eqnarray}
\end{subequations}
where  $\mathcal{N}_q$ is in GeV$^4$,  $\tilde{{\bf k}}_\perp = {\bf k}_\perp - (1-x) \Delta_\perp$, ${\cal M}_q^2(x)  =  x M^2 -x/(1-x) M_{X}^{q \, 2} -M_{\Lambda}^{q \, 2} $, $M$ is the proton mass, $m_q$ is the quark mass, $M_X^q$ the diquark system's mass (discussed below), and finally $M_\Lambda^q$ is the 
the mass term defining the coupling at the proton-quark-diquark vertex. Consistently with studies of baryons in the context of the application of Dyson--Schwinger equations in QCD (\cite{Roberts2} and references therein), we chose the coupling,  
\begin{equation}
\label{coupling}
\Gamma = g_s \frac{k^2-m_q^2}{(k^2- M_\Lambda^{q \, 2})^2},
\end{equation}
where $k^2$ is the four-momentum square at the vertex ($k^2 \rightarrow k'^2$ at the RHS vertex).
%
In order to construct a flavor dependent diquark model we start from the quark proton helicity amplitudes (Fig.\ref{fig2})), where the diquark can have spin $S=0,1$. Using  the SU(4) symmetry relations we construct the $u$ and $d$ components \cite{hybrid_even,Osvaldo_prep}. This step is described by the initial flowchart bubbles in Fig.\ref{fig:flow}.

The Regge term is given by
\begin{equation}
R^{\alpha_q,\alpha_q^\prime}_{p_q}=  x^{-[\alpha_q + \alpha_q^\prime(x) t  ]},
\label{regge}
\end{equation}
where  
\begin{equation}
\alpha_q^\prime(x) =  \alpha_q^\prime  (1-x)^{p_q}, 
\end{equation}
$\alpha_q'$ and $p_q$ being parameters. 

The nucleon Dirac and Pauli form factors were fitted with our model by considering, 
\begin{eqnarray}
\label{formf1}
F_1^{p(n)}(t) & = & \int dx H^{p(n)}(x,0,t) = e_{u(d)} \int dx H_{u}(t) +  e_{d(u)} \int dx H_{d}(t), \\
F_2^{p(n)}(t) & = & \int dx E^{p(n)}(x,0,t) = e_{u(d)} \int dx E_{u}(t) + e_{d(u)} \int dx E_{d}(t),
\end{eqnarray}
where $e_u=2/3$, and $e_d=-1/3$. 

In our recursive fitting procedure, we first set $t=0$, and constrained the Regge parameter,  $\alpha_q$, all of the mass parameters, and the normalization, by fitting 
the expression in Eq.(\ref{H}) to 
valence PDF distributions (this step includes perturbative QCD evolution).  We took the same set of parameters for the GPD $E$ since its forward limit cannot be constrained. 
By fixing these parameters, we obtain a total number of five per quark flavor, consistent with modern PDFs parametrizations (\cite{bench} and references therein). We then took the integrals of Eqs.(\ref{H},\ref{E}), which define the Dirac and Pauli form factors, respectively, and fitted the remaining parameters, $\alpha'_q$ and $p_q$ in the Regge term in Eq.(\ref{regge}), to 
the available proton and neutron form factors \cite{GEP_exp,GMP_exp,GEN_exp,GMN_exp,GEP_GMP}. 

\begin{figure}
\includegraphics[width=7.0cm]{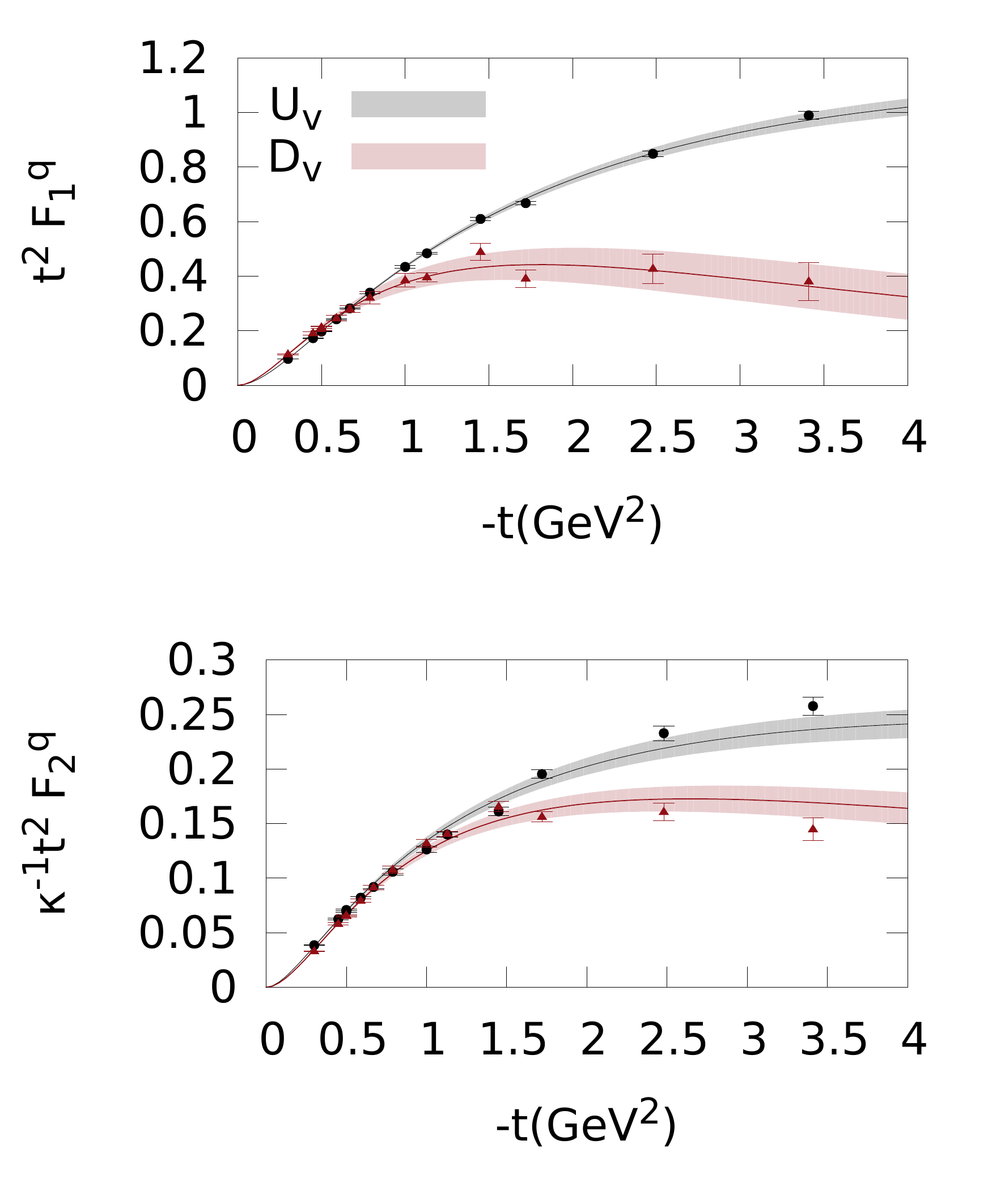}
\includegraphics[width=7.0cm]{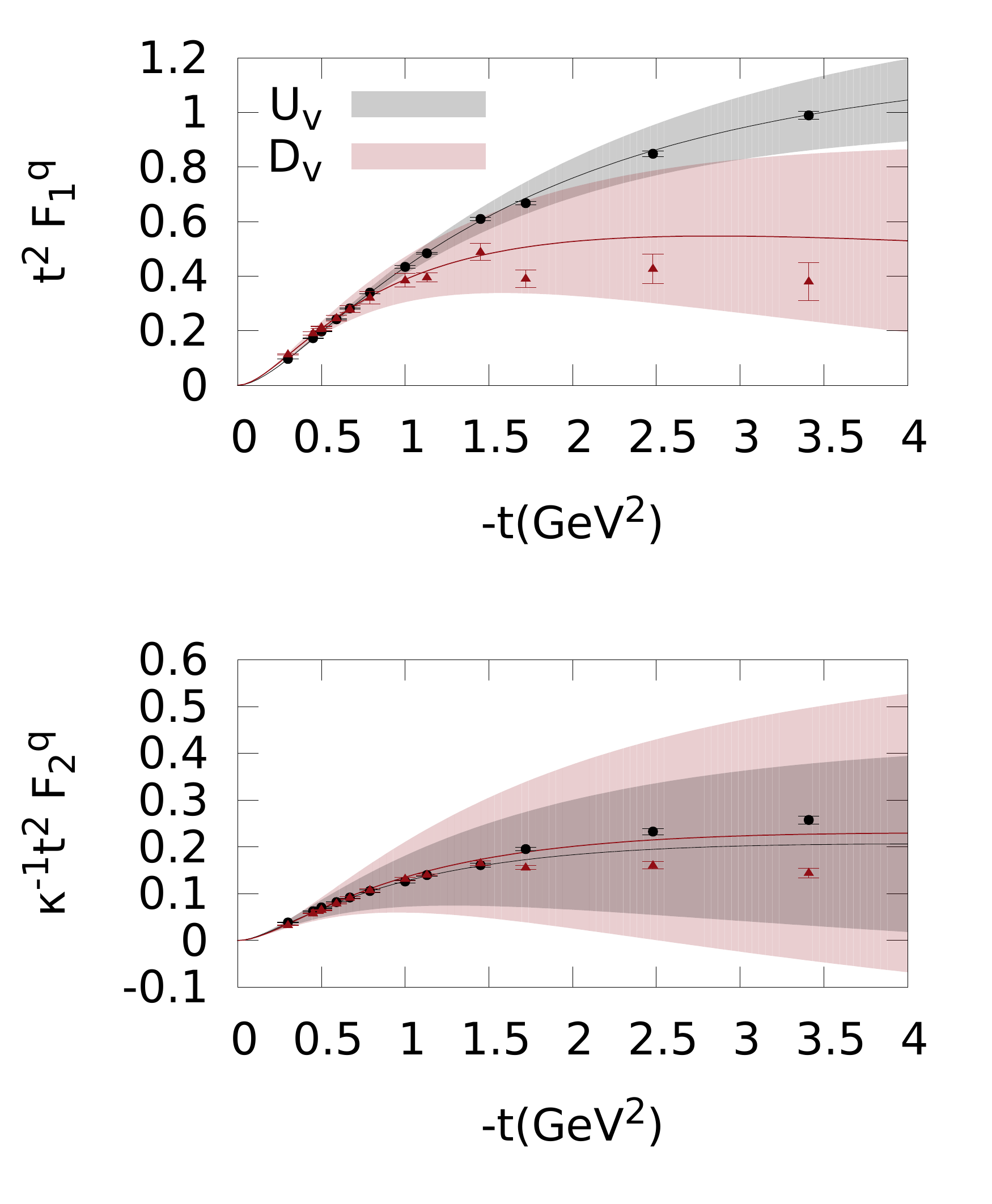}
\caption{(Color online) Right Panel: Proton form factors $t^2 F_1^q$ (top) and $\kappa_q^{-1} t^2 F_2^q$ (bottom) plotted vs. $-t$, as obtained from our parametrization using  Eq.(\ref{fit_form}), by fitting the data preceding Ref.\cite{Cates}. Left Panel: same as Right, but including the data from Ref.\cite{Cates}. The parametrization's error bands are from our GPD fit, and they reflect the errors on the parameters displayed in Table \ref{table1}, obtained without taking into account parameters' correlations. $\kappa_u = 2.03$, and $\kappa_d = -1.67$, are derived in terms of  the proton and neutron anomalous magnetic moments using isospin symmetry (cf. Eq.(\protect\ref{iso}) at $t=0$). Experimental data on both panels are from Ref.\cite{Cates}.}
\label{fig4}
\end{figure}
Using the new form factor data \cite{Cates} to constrain $\alpha'_q$ and $p_q$, we repeated  the fit  keeping the PDFs parameters fixed (central bubbles in Fig.\ref{fig:flow}). 
n Ref.\cite{Cates} the $u$ and $d$ quarks contributions to the nucleon form factors were extracted up to values of $t$ in the  few GeV region, using  
the isospin symmetry decomposition formulae,
\begin{subequations}
\label{iso}
\begin{eqnarray}2
F_{1(2)}^u & = & 2 F_{1(2)}^p + F_{1(2)}^n \\  
F_{1(2)}^d & = & 2 F_{1(2)}^n + F_{1(2)}^p 
\end{eqnarray}
\end{subequations}   
where, as usual, $F_{1(2)}^u$, and $F_{1(2)}^d$ are the Dirac and Pauli $u$ and $d$ quarks contributions to the proton form factor. 
\footnote 
{It should be also noticed that both trends appear by extrapolating  the widely used form factors  parametrization by Kelly \cite{Kelly} to large $t$.}
%

In Table \ref{table1} we show the values of $\alpha'_q$ and $p_q$, obtained using both the old set of data and the data from Ref.\cite{Cates}.
\begin{table}[h]
\begin{tabular}{|c|c|c|c|c|}
\hline
\hline

Parameters             &  $H$  old data     & $H$ Ref.\cite{Cates}  & $E$ old data        & $E $  Ref.\cite{Cates}\\ 
\hline
\hline
$\alpha^\prime_u$      & 1.889 $\pm$ 0.0845  & 1.814 $\pm$ 0.0220    &  2.811 $\pm$ 0.765  &     2.835 $\pm$ 0.0509\\
$p_u$                            & 0.551 $\pm$ 0.0893  & 0.449 $\pm$ 0.0170    &  0.863 $\pm$ 0.482  &     0.969 $\pm$ 0.0307\\
$ \chi^2$              & 0.8                & 0.9                   &  0.7                &     4.8              \\
\hline
$\alpha^\prime_d$      & 1.380 $\pm$ 0.145  &  1.139 $\pm$ 0.0564    &  1.362  $\pm$ 0.585 &   1.281 $\pm$ 0.0310  \\
$p_d$                            & 0.345 $\pm$ 0.370  & -0.113 $\pm$ 0.104      &  1.115  $\pm$ 1.150 &   0.726 $\pm$ 0.0631  \\
$ \chi^2$              & 0.8                & 0.5                   &  0.7                &    3.2               \\
\hline

\end{tabular}
\caption{%
\label{table1} 
Parameters obtained from our recursive fitting procedure applied to $H_q$, $E_q$,  $q=u,d$.  
We obtained  $\alpha^\prime_q$, $p_q$, by fitting the proton and neutron electromagnetic form factors. Also shown are the $\chi^2$ values for the separate contributions to the fit.}
\end{table} 
The error analysis was performed using the Hessian method, and we could therefore check the effect of correlations between parameters. 
Since we fix the values of the PDFs parameters, only the form factor parameters, $p_q$, and $\alpha'_q$ will show the effect of correlations. We do find some degree of correlation between these two parameters, that we plan to study in more detail in future work including more flexible parametrizations. In order to perform this type of analysis in a fully quantitative way it will, however, be necessary to include also experimental error correlations, which do not exist in most cases in published form.  
We would like to point out that these type of observations are now possible exactly because of the precision attained using the flavor separated data. 
The new fit displayed in Fig.\ref{fig4} allowed us, in fact, to reduce the errors on the GPD parameters considerably.
Notice that both the $x$ and $t$ dependences of the GPDs are affected by the change in $\alpha'_q$ and $p_q$. Changes occur in the $x$ shapes of all four GPDs, $H_{u,d}$ and $E_{u,d}$, the GPDs largest variations being at large $t$ and small $x$.  

In summary, our model is given by the expressions in Eqs.(\ref{fit_form},\ref{H},\ref{E},\ref{regge}) and Ref.\cite{hybrid_even}. While the mass parameters and the normalizations have been kept fixed to the values given in Ref.\cite{hybrid_even}, the  parameters  $\alpha^\prime_q$, $p_q$ have been re-fitted using the new flavor separated data from Ref.\cite{Cates}, and are listed in Table \ref{table1}. 

\subsection{Reggeization}
Although the expression in Eq.(\ref{fit_form}) was purposely cast in a simple enough form to be used in fits of data  \cite{AHLT1,AHLT2,hybrid_even}, it can be considered a phenomenological realization of a more 
elaborate model including diquark correlations through Regge duality. 
As we explain below, $\alpha$ defines a Regge trajectory, while $\alpha'$ and $p$ regulate the effect of  Regge cuts, which in turn can be interpreted as diquark correlations. 

\subsubsection{Regge Factorization}
\begin{figure}
\includegraphics[width=11.0cm]{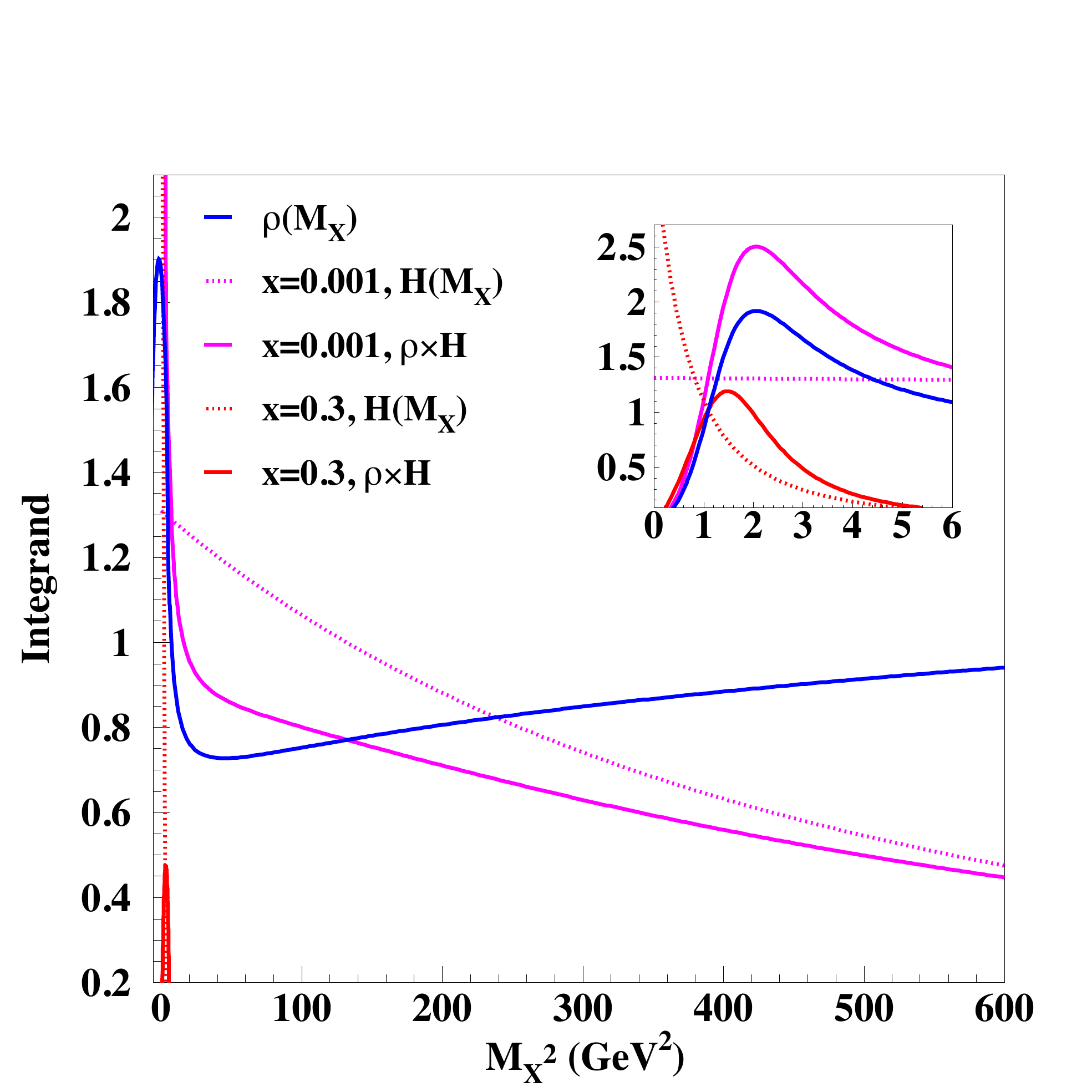}
\caption{(Color Online) Integrand components in Eq.(\ref{reggeization}) plotted vs. $M_X^2$. The various curves show the spectral function, $\rho(M_X^2)$, defining the reggeization of the quark-proton scattering amplitude,  the contribution of $H_{M_X,m}^{M_\Lambda}(x,0,0)$  at two values of  $x$, $x=10^{-3}$ and $x=0.3$, and  the product of $\rho(M_X^2)$ and  $H_{M_X,m}^{M_\Lambda}(x,0,0)$, Eq.(\protect\ref{reggeization}) for the same values of  $x$.   The inset highlights the low mass region.}
\label{fig5}
\end{figure}

In order to explain our model, and specifically  the role of the Regge parameters  $\alpha$, $\alpha'$ and $p$
\footnote{
In this section we drop the $q$ subscript for simplicity} we 
first illustrate the reggeization procedure. 
We consider the spin independent GPD, $H$, in the forward limit, $H(x,0,0)=f_1(x)$, for simplicity,  
as a function of a continuum of diquark masses. We follow the procedure first outlined in Ref.\cite{BroCloGun}, although we extend it to all $M_X$, including low values of the mass. 
The exact expression obtained using a fixed mass scalar diquark
is given by,
 \begin{eqnarray}
 H_{M_X,m}^{M_\Lambda}(x,0,0)  & = &
{\cal N} \frac{\pi}{12}  \left[\frac{2(m + x M)^2+ {\cal M}^2(x, M_\Lambda^2,M_X^2)}{{\cal M}^6(x, M_\Lambda^2,M_X^2)} \right]
 (1-x)^4, 
 \label{HX00}
\end{eqnarray}
where
\[ {\cal M}^2(x, M_\Lambda^2,M_X^2) = x M^2 - \frac{x}{1-x} M_X^2 - M_\Lambda^2. \] 
We multiply this expression with a spectral density $\rho_R(M_X^2)$ of the type shown in Fig.\ref{fig5}.  
The spectral function was parametrized as, 
\begin{equation}
\label{rho}
 \rho(M_X^2) = \frac{(M_X^2)^a}{[1+b(M_X - \overline{M}_X)^2]^c} \equiv (M_X^2)^a B(M_X), 
\end{equation}
where $c=a+1-\alpha$, and the quantity $M_X^2 = M_X^2/(1 \, {\rm GeV^2})$ is dimensionless. Schematically, we summarize the behavior of $\rho(M_X^2)$ as,
\begin{eqnarray}
\label{rho_approx}
 \rho(M_X^2) \approx  \left\{  
 \begin{array}{crrl}  (M_X^2)^{\alpha-1} & & M_X^2 &  \rightarrow \infty \\
 & \\ 
 \delta(M_X^2-\overline {M}_X^2) & & M_X^2 &  {\rm few \; GeV^2}
 \end{array}  \right.
\end{eqnarray}
By integrating over $M_X^2$  one obtains that the Regge behavior in $x$ factors out, namely,
\begin{subequations}
\label{reggeization}
\begin{eqnarray}
H(x,0,0) & = & {\cal N}  \int_0^\infty  d M_X^2  \rho(M_X^2)  H_{M_X,m}^{M_\Lambda}(x,0,0) \label{line1} \\
& = &  {\cal N}   x^{-\alpha} \int_0^\infty d z \,  z^{\alpha-1}  B(z/x) H_{M_X,m}^{M_\Lambda}(x,0,0)  \label{line2} \\
& \approx &   {\cal N} x^{-\alpha}  H_{\overline{M}_X,m}^{M_\Lambda}(x,0,0)  .
\label{line3}
\end{eqnarray}
\end{subequations}
where $z= X M_X^2$. 
In Fig.\ref{fig5} we also show the integrand in Eq.(\ref{reggeization}) as a function of $M_X^2$ (the inset was drawn to highlight the low masses behavior).
One can see the different behavior for small and large values of $x$, the large $x$ behavior being characterized by a flatter slope in $M_X$. 
Notice also that the integrands peak at low $M_X$ (inset Fig.\ref{fig5}), the position of the peak being fixed in a common mass range $1 \lesssim M_X^2 \lesssim 2.5 $ GeV$^2$, 
despite the wide variation in $x$ values.  
As a consequence, once the Regge behavior is factored out as in the second line of Eq.(\ref{reggeization}),  an integration over an almost  ``$\delta$-like" peak remains, yielding the last line of Eq.(\ref{reggeization}). 
In other words, the average of the weighted integrand over the full range of $z$ in Eq.(\ref{line1}) is approximately the integrand's value at the peak of the spectral distribution in Eq.~(\ref{line3}).
Switching on the skewness and $t$ dependences in Eq.(\ref{reggeization}), one obtains the result in Eq.(\ref{fit_form}), where Regge behavior is cast in a  factorized (Regge times diquark) form.

\begin{figure}
\includegraphics[width=8.5cm]{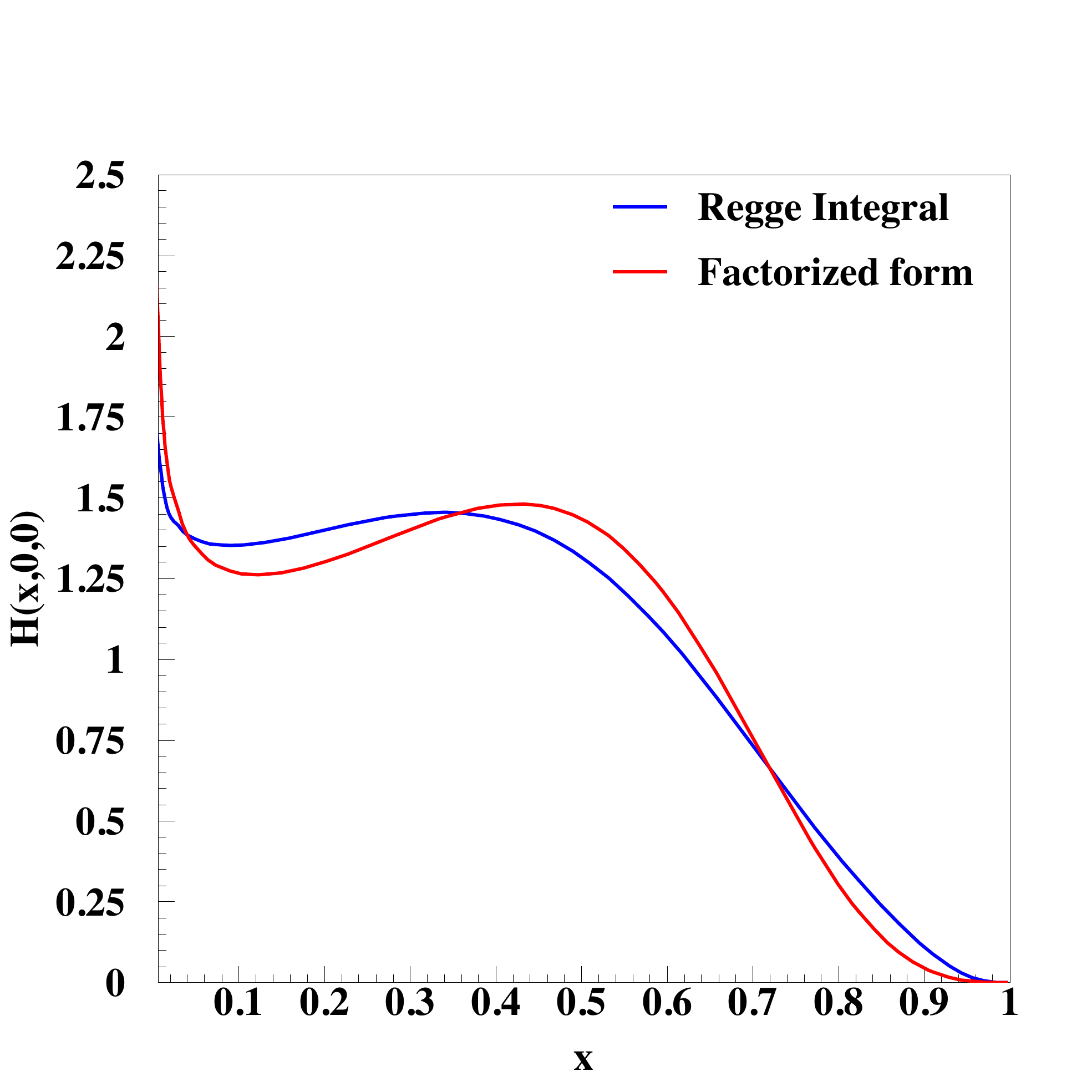}
\caption{(Color Online) Comparison between $H(x,0,0)$ calculated using the factorized ansatz in Eq.(\ref{fit_form}), and the full expression in Eq.(\ref{reggeization}).}
\label{fig6}
\end{figure}

The function $H$ in Eq.(\ref{reggeization}), obtained using both the factorized ansatz (Eq.~\ref{line3}) and the full calculation, is shown in Fig.\ref{fig6}. 
The figure demonstrates that the factorized form of Eq.(\ref{fit_form}) can be taken for all values of $x$, thus extending the validity of the original model of Ref.\cite{BroCloGun}, in the hypothesis that an appropriate form for the recoiling system's invariant mass spectral function is considered. 
In the diquark model part of Eq.(\ref{fit_form}), $H_{{M}_X,m}^{M_\Lambda}(X,0,t)$, $M_X \approx \overline{M}_X$, from Eqs.(\ref{rho},\ref{rho_approx}). 

It is important to note that this model is in line with most forms used for PDFs parameterizations, which also display a factorized Regge term.  
The fact that the Regge term can be factorized for all values of $x$ does not imply that Regge behavior can be extended to all values of $x$. The 
$x$ dependence away from small $x$ is indeed quite different (see discussion in \cite{AHLT1}). Also, the values of the Regge parameters obtained from our fits do not coincide exactly with known Regge trajectories owing to the fact that the diquark model provides a ``tail" at small $x$ that contributes to the slope. This point was extensively discussed in Ref.\cite{AHLT1} (see Fig.12 in \cite{AHLT1}).

\subsubsection{Interpretation}
\begin{figure}
\includegraphics[width=9.0cm]{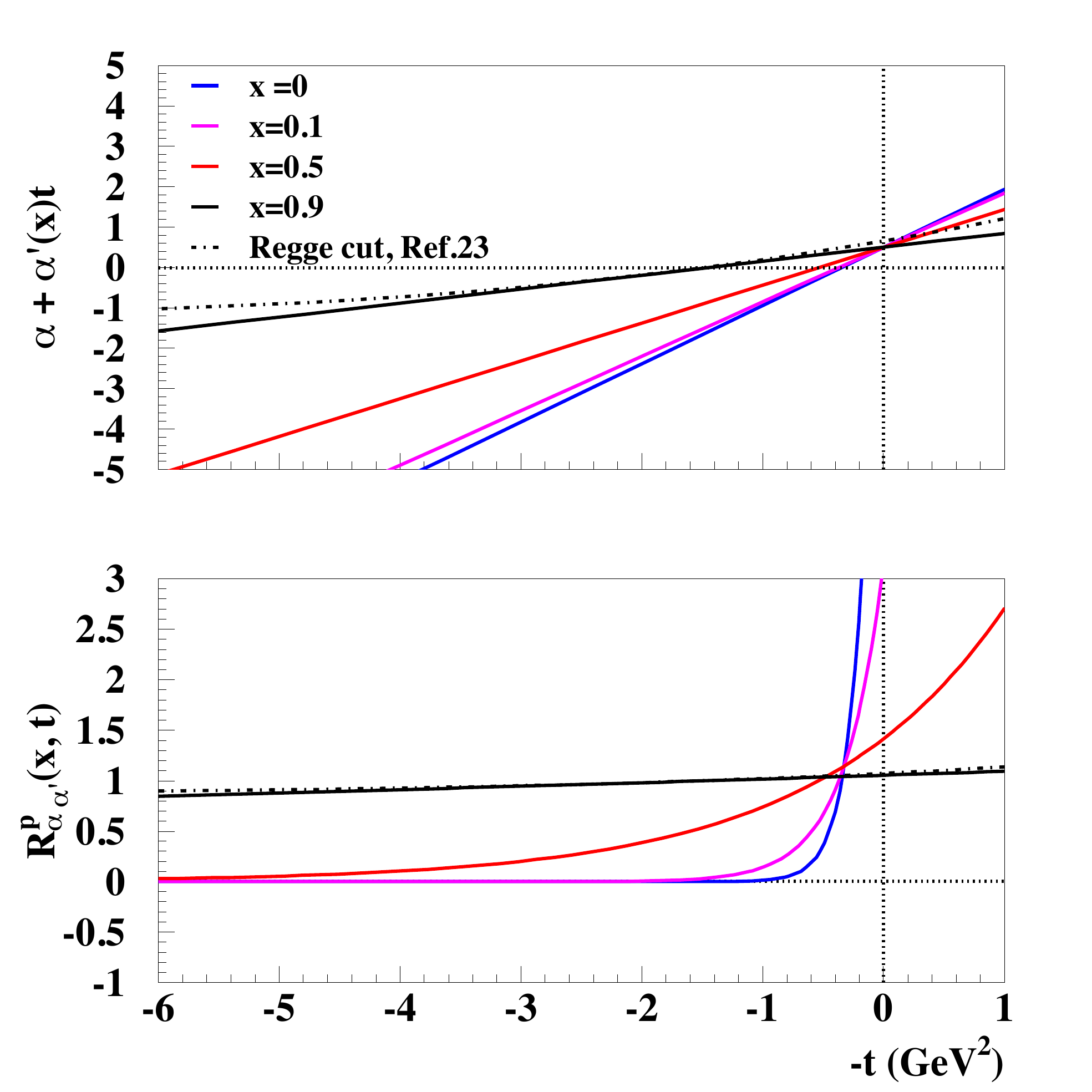}
\caption{(Color online) Combined $x$ and $t$ dependences of the Regge contribution in our model (Eq.(\ref{regge}) and discussion in text). Upper panel: trajectories with $x$-dependent parameters 
from our model. The dot dashed line is a comparison with the softening of the slope first suggested in Refs.\cite{ColKea,PDBCol}.  Lower panel: Regge term, Eq.(\ref{regge}) obtained using the trajectories shown in the upper panel.
}
\label{fig7}
\end{figure}

The Regge term has different interpretations depending on the kinematical regions considered. The different regions are displayed in Fig.\ref{fig7} where in the upper panel we show trajectory,  $\alpha + \alpha^\prime(x) \, t$, plotted vs. $t$ for different values of $x$, while in the lower panel we show the Regge term, Eq.(\ref{regge}), correspondingly  obtained using the trajectories from the upper panel.
In our calculation we consider two main regions:

\noindent \underline{ $\xi=\zeta=0$, $t \rightarrow 0$}  

\noindent 
Eq.(\ref{reggeization}) can be generalized straightforwardly from $t=0$ to arbitrary small $t < 0$, noticing that  the integrals are done at fixed $t$. So long as $\alpha +  \alpha^\prime (1-x)^p \, t$ remains positive, {\it i.e.} for $-t<\alpha/\alpha'(x)$,  the physical interpretation remains unaltered because the behavior of the $M_X^2$ spectrum does not change considerably (this occurs for $t \lesssim 0.5$ GeV$^2$, see Fig.\ref{fig4}(a)). 
This is the region spanned in DVCS-type experiments, where the reaction's four-momentum transfer $Q^2 \approx$ (few GeV$^2$), and $-t < 1$ GeV$^2$. 

\vspace{0.3cm}
\noindent \underline{ $\xi=\zeta=0$, $t \neq 0$, $t<<s,u$} 

\noindent Larger, but not asymptotic,  negative $t$ can still be consistent with Regge behavior.
In fact, the $x$ dependence of $ \alpha^\prime(x)$ becomes important in this kinematical region. This is a consequence of Regge cuts \cite{ColKea,PDBCol}. 
More specifically, the linear Regge trajectory gets modified due to multiple interactions. The first interaction is given by the exchange of two pomerons which
modifies the slope in $x^{-(\alpha+\alpha' t)}$
 because of the presence of the second pomeron.\footnote{
As a reminder $\hat{s}^{(\alpha+\alpha' t)} \rightarrow x^{-(\alpha+\alpha' t)}$. } The next interaction modifies the slope a bit more. The following multiple interactions eventually produce the flattening of the slope. This is used to extend the theory far from $t=0$.
%

In summary, the $x$-dependence of the slope parameter, $\alpha +  \alpha'(x) t $, is such that it softens the decreasing behavior of the trajectory with negative $t$ with respect to the standard Regge behavior described by $\alpha +  \alpha'(0) t $ (blue curve in the figure). 
The behavior with $t$ is illustrated in Fig.\ref{fig4}a where one can see that the slope in $t$ of the exponent, $\alpha+\alpha' (x) t$, decreases with $x$ eventually flattening out. The dot-dashed curve was obtained in a calculation from Ref.\cite{ColKea,PDBCol} including the effect of multiple interactions/Regge cuts which is well reproduced by our form. 

\begin{figure}
\large{\bf (a)}
\includegraphics[width=6.0cm,angle=0]{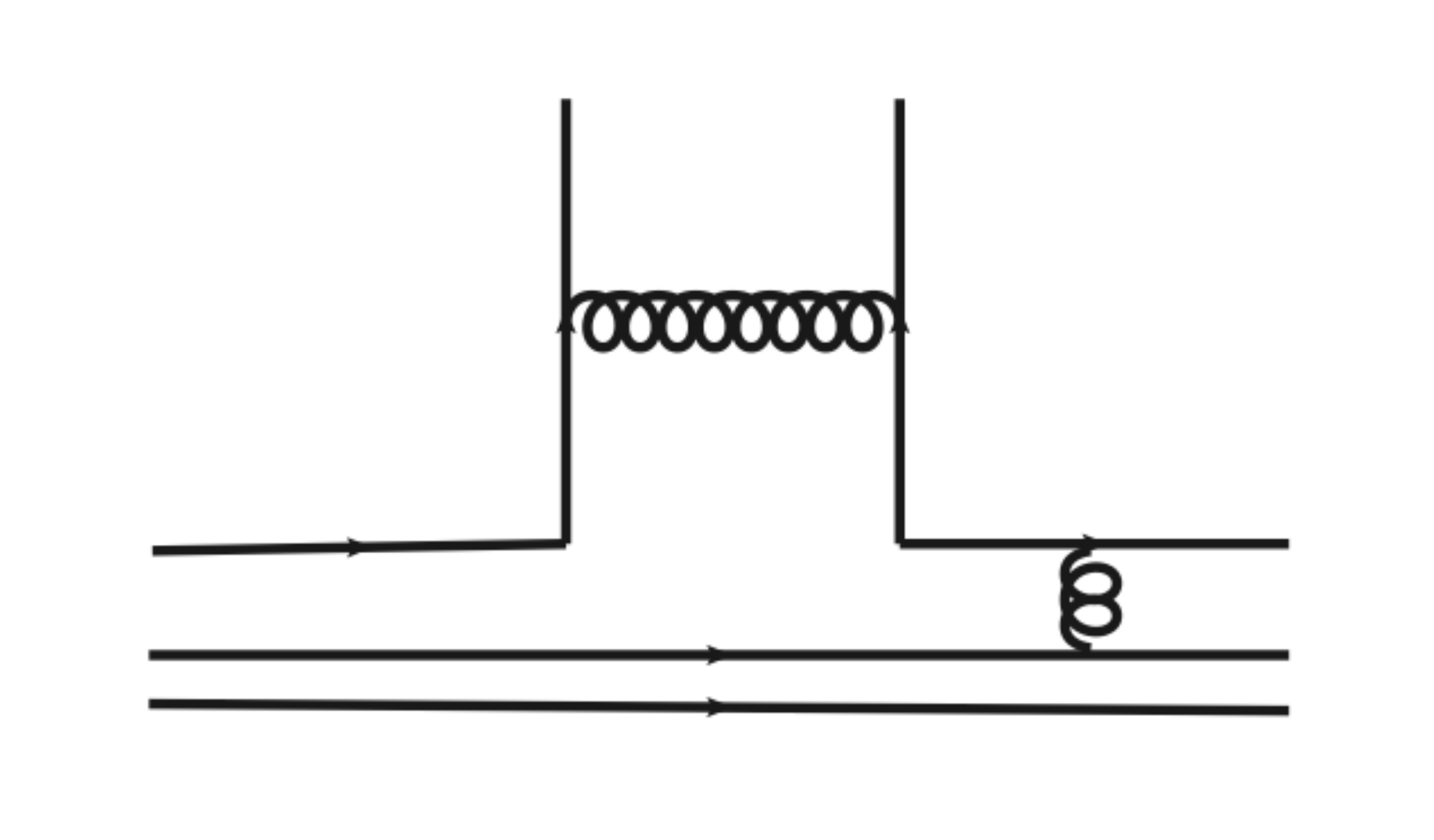}
\hspace{0.5cm}
\large{\bf (b)}
\includegraphics[width=6.0cm,angle=0]{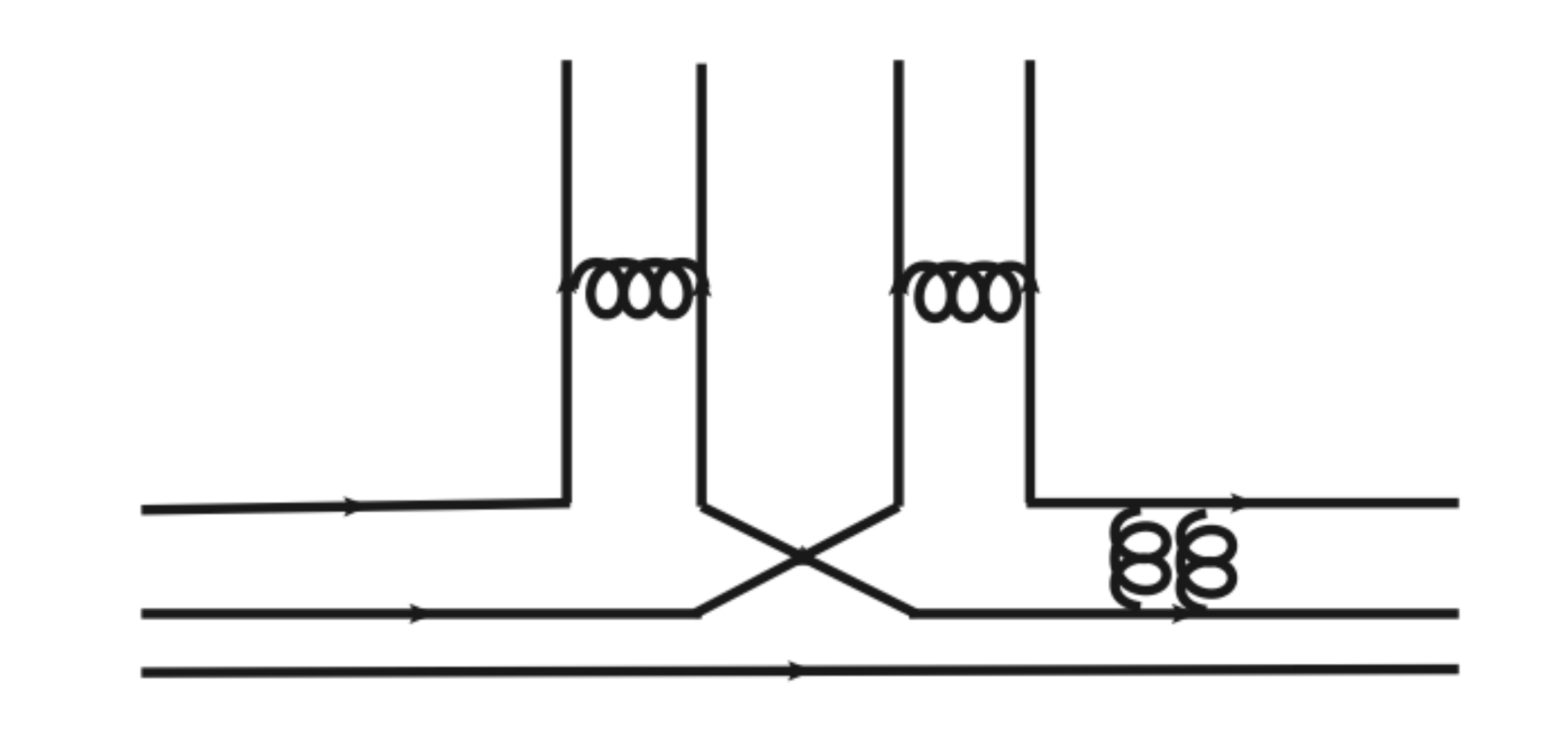}
\normalsize
\caption{Parton content of Regge exchanges. (a)  Regge exchange with no interactions (Fig.\ref{fig4}a); (b) Regge cut with partonic structure corresponding to a diquark correlation.}
\label{fig8}
\end{figure}

The partonic structure that gives rise to the Regge behavior discussed so far is represented in Fig.\ref{fig8}. 
Fig.\ref{fig8}a describes a simple Reggeon exchange while Fig.\ref{fig8}b exhibits a Regge cut, or a reinteraction. 
Notice that the non-planar structure of this graph allows us to interpret the Regge behavior in this region as given by coupling of diquarks to the virtual photon.
Our proposed picture therefore connects with the one considered in the Poincar{\'e} covariant Dyson-Schwinger equation (DSE)  approach of Ref.\cite{Roberts,Roberts3}, including both 
dressed quark and diquark components coupling to the virtual photon.

Finally, we note that including  skewness, {\it i.e.} $(\xi, \zeta) \neq 0$, is more complicated, as singularities at the crossover point, $X=\zeta (x=\xi)$, might arise.
This issue has been extensively discussed recently in Ref.\cite{Rad11} in the context of Double Distributions (see also \cite{Szcz11}), although in the $t=0$ limit. 
Our results in this paper, however, do not depend on the $\zeta \neq 0$ case: we interpret the form factors using the zero skewness section of GPDs, which allows for a clear interpretation in $b$-space, as we explain in what follows. 

In conclusion, with a viable model in hand we can now examine the role and the interplay of its different components in interpreting a variety of experimental data.  
In particular, since the $t$ dependence arises naturally in our model, namely it is not superimposed {\it ad hoc},  we can understand what features of the GPDs can  simultaneously fit the PDFs and feed into the form factors.
Why can, for instance, our flexible model reproduce the flavor dependence of the nucleon form factors? Which components are dominant -- Regge or diquark, and for what values of t? What aspect of the nucleon substructure can this be traced back to? 
    
\section{Interpretation of Flavor Dependence}
\label{sec:3}
In order to study the interplay between single quark scattering and the reinteraction terms (Regge cuts), and to ascertain whether either component can by itself drive  the $t$-dependence,  we define starting from Eqs.(\ref{fit_form}) to (\ref{regge}), the following  additive form, 
\begin{eqnarray}
\label{separation1}
H_q(X,\zeta,t)  = H_{diq+R}^q + H_R^q,  \;\;\;\;\;\;  E_q(X,\zeta,t)  = E_{diq+R}^q + E_R^q
\end{eqnarray}
where,
\begin{eqnarray}
\label{RG1}
H_{diq+R}^q & = &  {\cal N}_{diq+R}^q \, H_{M_X^q,m_q}^{M_\Lambda^q} R^{\alpha_q}_{H, p_q}, \;\;\;\;\;\;  
E_{diq+R}^q   =    {\cal N}_{diq+R}^q \, E_{M_X^q,m_q}^{M_\Lambda^q} R^{\alpha_q}_{E, p_q}  \\
\label{RG2}
H_R^q & = & {\cal N}_{diq+R}^q \, R^{\alpha'_q}_{H, p_q}, \;\;\;\;\;\; E_R^q  =  {\cal N}_{diq+R}^q \, R^{\alpha_q}_{E, p_q}.
\displaystyle{}  
\end{eqnarray}
with 
\begin{equation}
{\cal N}_{diq+R}^q = \left( 1/R^{\alpha_q,\alpha^\prime_q}_{H(E), p_q} +  1/H(E)_{M_X^q,m_q}^{M_\Lambda^q} \right)^{-1}.
\end{equation}
%
Notice that in the Regge term Eq.(\ref{regge}), we separated out the non-interacting, and interacting components (Figs.\ref{fig7}, \ref{fig8}), which are defined from ,
\begin{equation}
R^{\alpha_q, \alpha'_q}_{H(E), p_q}(x,t) = R^{\alpha_q}_{H, p_q}(x)  \, R^{\alpha'_q}_{H(E), p_q}(x,t).
\end{equation}

Using Eqs.(\ref{RG1},\ref{RG2}), the flavor dependent Dirac and Pauli form factors can respectively be written as,
 \begin{eqnarray}
 F_1^q & = &  \int_0^1 dx (H_{diq+R}^q+H_{R}^q), \\
 F_2^q & = &  \int _0^1 dx (E_{diq+R}^q+E_{R}^q), 
 \end{eqnarray}

\begin{figure}
\includegraphics[width=8.0cm]{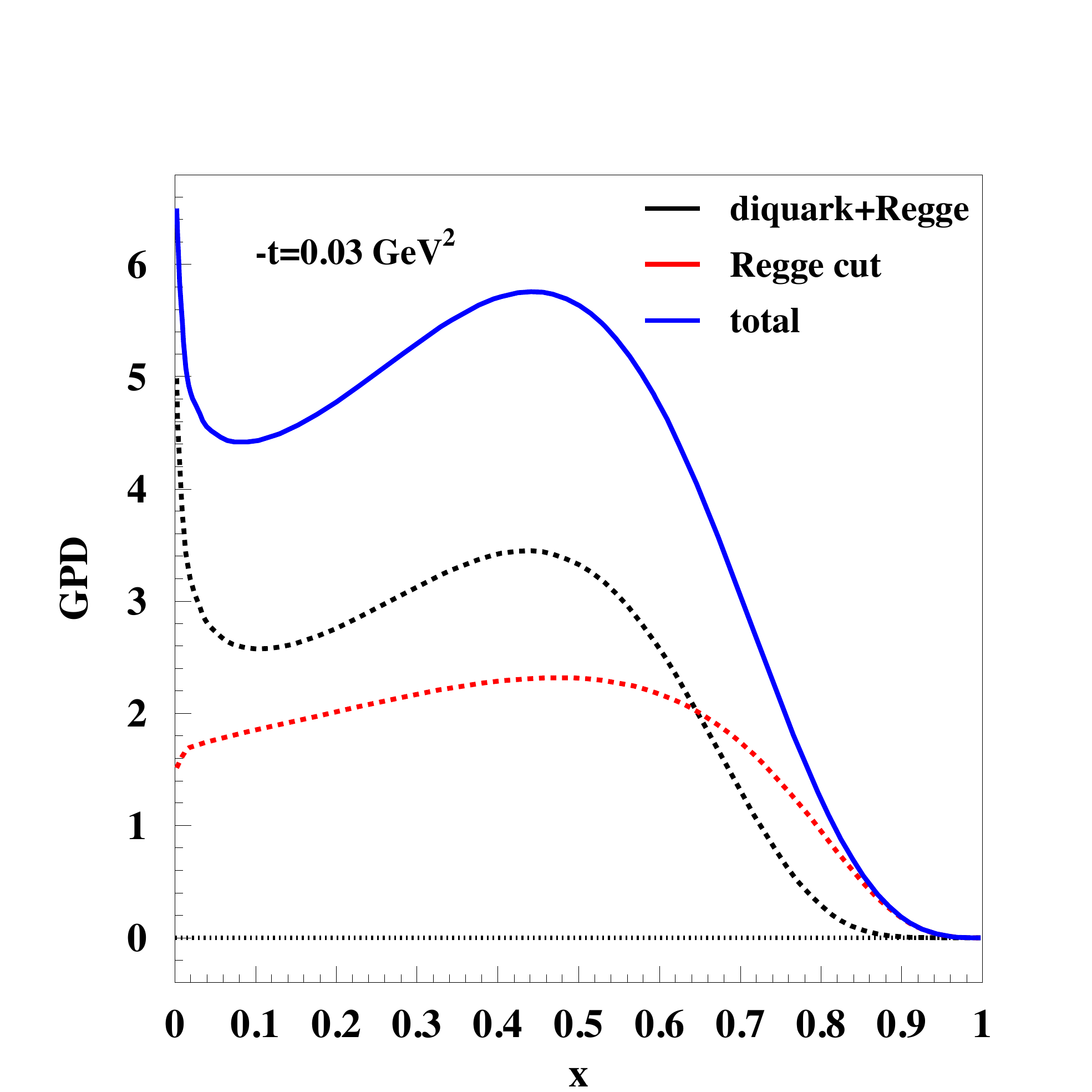}
\includegraphics[width=8.0cm]{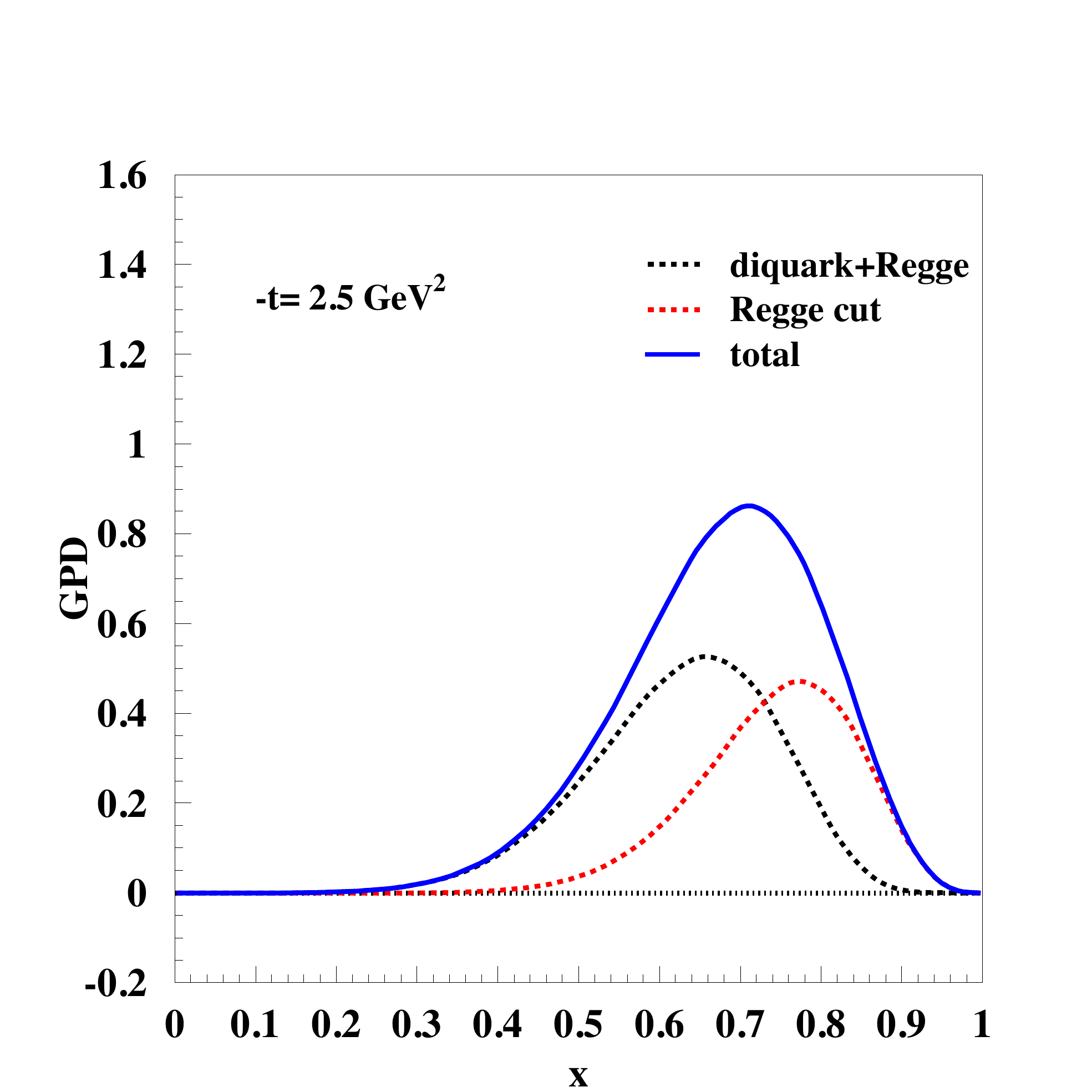}
\includegraphics[width=8.0cm]{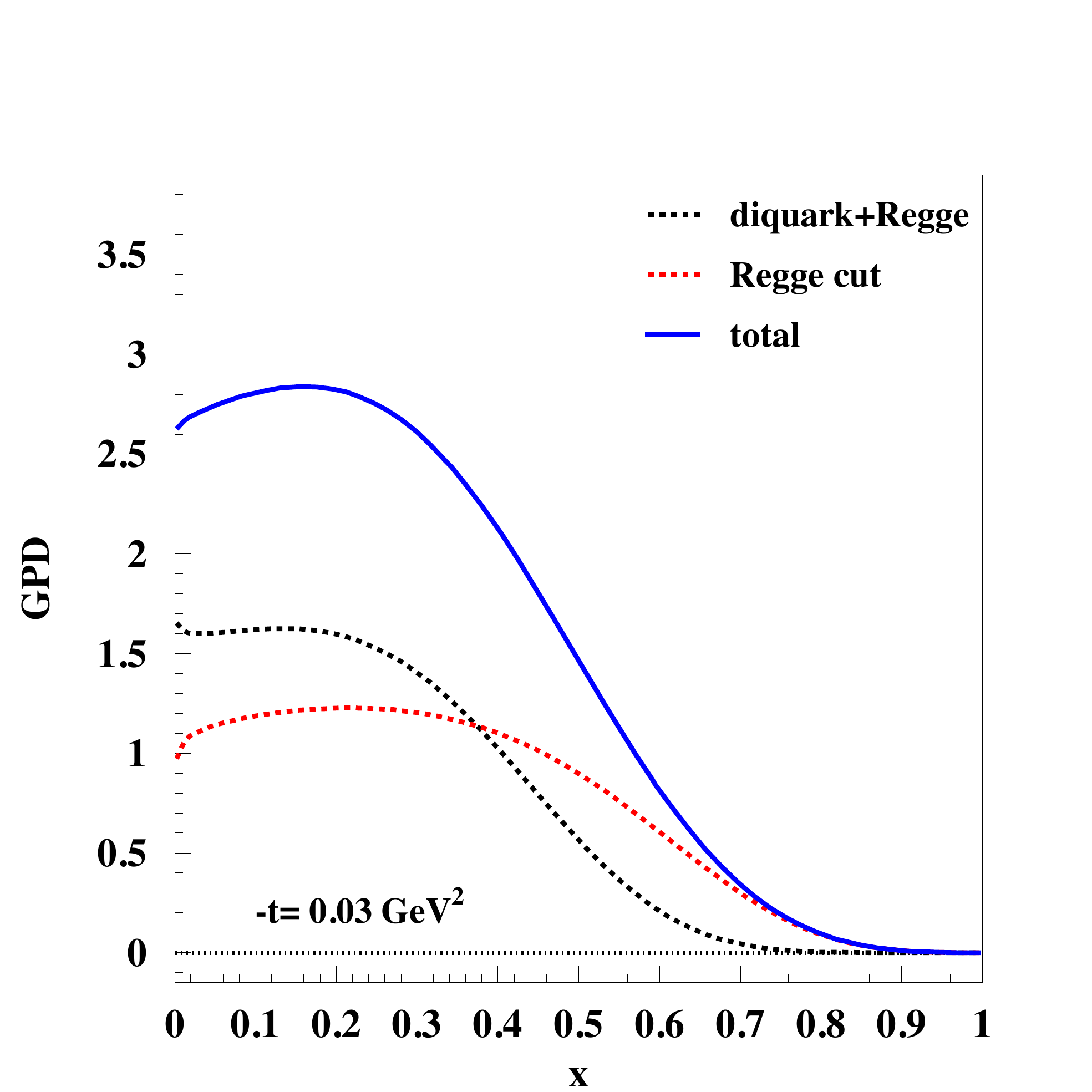}
\includegraphics[width=8.0cm]{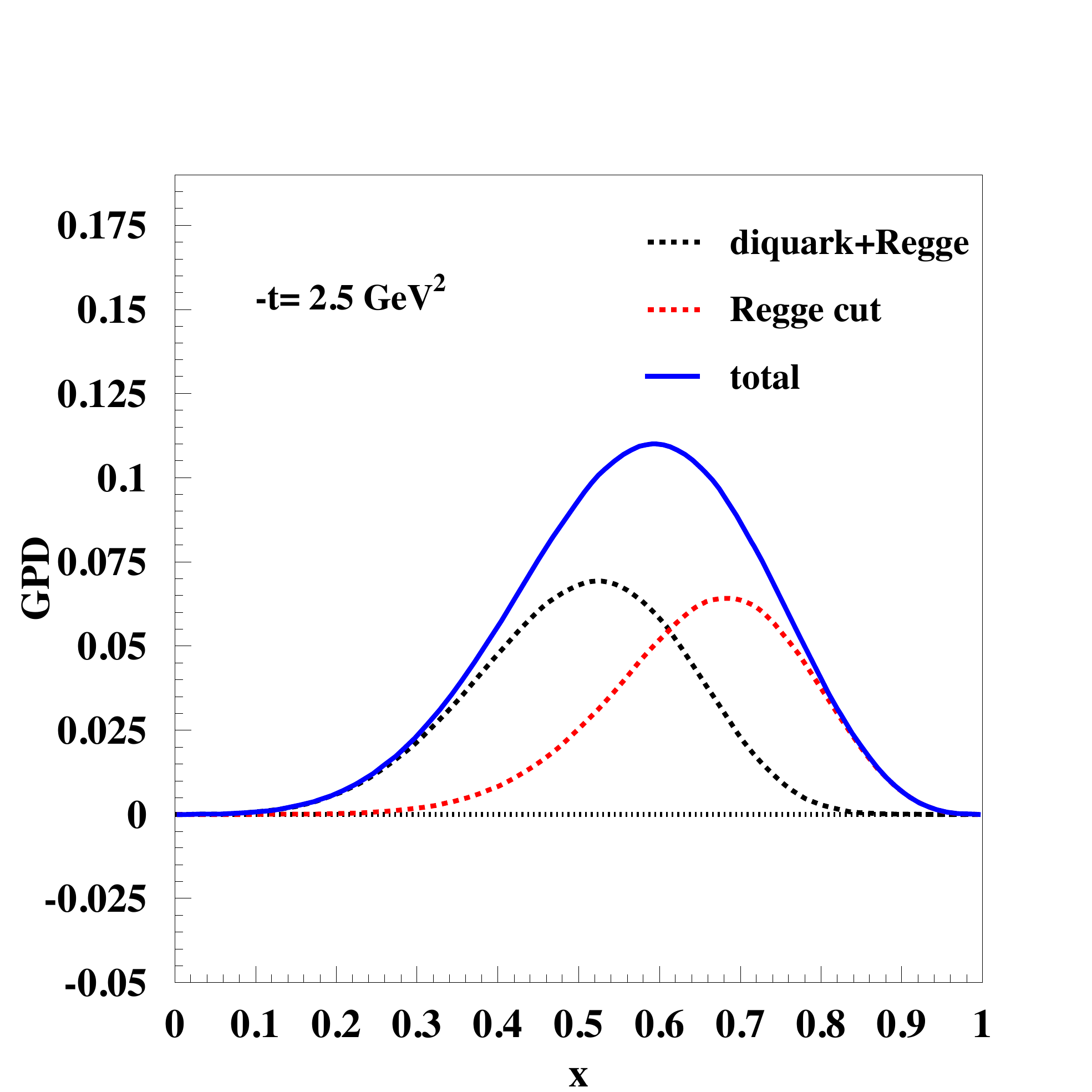}
\caption{(Color online) Components of the GPD $H_u$ (upper panels) and $H_d$ (lower panels) contributing to the form factor, $F_1^q$, $q=u,d$, for $-t =0.03$ GeV$^2$ (left), and $-t = 2.5$ GeV$^2$ (right). 
Separately displayed are the diquark plus $t$-independent Regge contribution, and the $t$-dependent Regge contribution which effectively takes into account Regge cuts, or diquark correlations (Figs.\ref{fig7} and \ref{fig8}).  }
\label{fig9}
\end{figure}
\begin{figure}
\includegraphics[width=8.5cm]{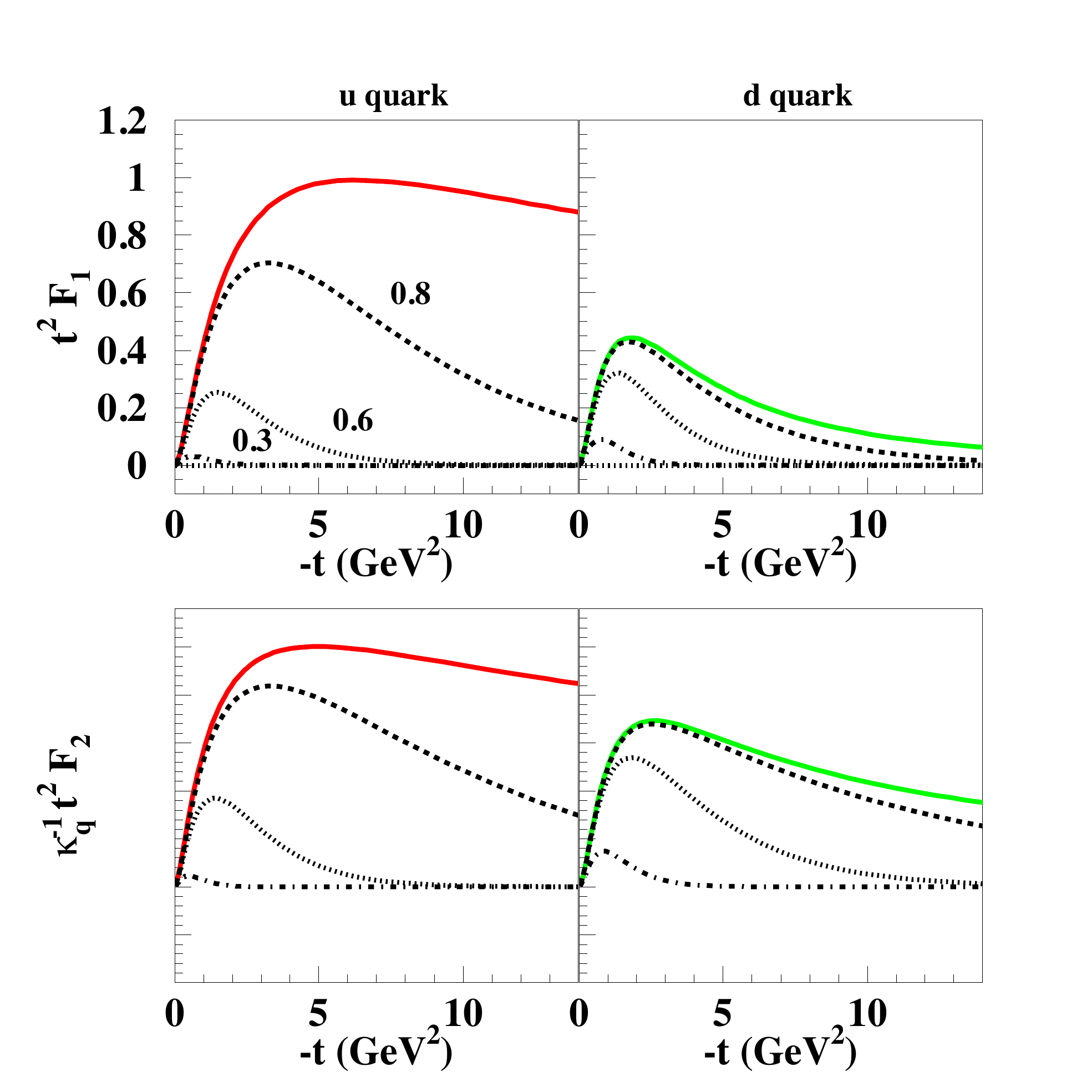}
\caption{(Color online) Contribution of various $x$ components to the proton form factors $t^2 F_1^q$ (top) and $\kappa_q^{-1} t^2 F_2^q$ (bottom) plotted vs. $-t$, obtained from our parametrization (Fig.\ref{fig7}). The u quarks are plotted to the left and the $d$ quarks to the right. The different lines correspond to $x_{MAX} = 0.3, 0.6,  0.8$ as indicated on the upper left figure. }
\label{fig10}
\end{figure}
In what follows we give an interpretation of the behavior seen in \cite{Cates}  that at the largest $t$ of the experiment 
the contribution of the $d$ quark is suppressed relative to $u$, for both $F_1$ and $F_2$.

\subsection{Momentum Space Analysis}
In Fig.\ref{fig9} we show the components of the GPD $H_q$ contributing to the form factor, $F_1^q$, $q=u,d$, for $t = -0.03$ GeV$^2$, and $t = -2.5$ GeV$^2$. 
We display separately the diquark plus $t$-independent Regge contribution, and the $t$-dependent Regge contribution which effectively takes into account Regge cuts, or diquark correlations, Eqs.(\ref{RG1},\ref{RG2}).
One can see that for both the $u$ and $d$ quarks, diquark correlations dominate over the diquark component at large $x$. 
Notice that at large $t$  the $u$ quarks behave differently from the $d$ quarks in that both the single quark scattering and the diquark correlation contributions are shifted to higher values of $x$ for the $u$ quarks. In other words, the $u$ quarks are governed by higher $x$ components. In order to test the consequence of this feature on the form factors flavor dependence, in Fig.\ref{fig10} we show,
\begin{eqnarray}
 t^2 F_1^{q, x_{MAX}} & = &   t^2  \int_0^{x_{MAX}} dx (H_{diq}^q+H_{R}^q), \\
\kappa_q^{-1} t^2  F_2^{q, x_{MAX}} & = & \kappa_q^{-1} t^2  \int _0^{x_{MAX}} dx (E_{diq}^q+E_{R}^q), 
 \end{eqnarray}
for different values of  $x_{MAX}<1$. The values of $x_{MAX}$ reported in the figure are: $x_{MAX} = 0.3, 0.6,  0.8$. One can see that the d quark form factor saturates at smaller $x_{MAX}$, as expected from Fig.\ref{fig10}. Therefore, we conclude that for  both the $u$ and $d$ quark, integrating over the peak given by diquark correlations (second peak in Fig.\ref{fig10}) is important. The form factors' behavior at $t$ in the multi-GeV region is governed in our model by re-interactions.    

\begin{figure}
\includegraphics[width=9.5cm]{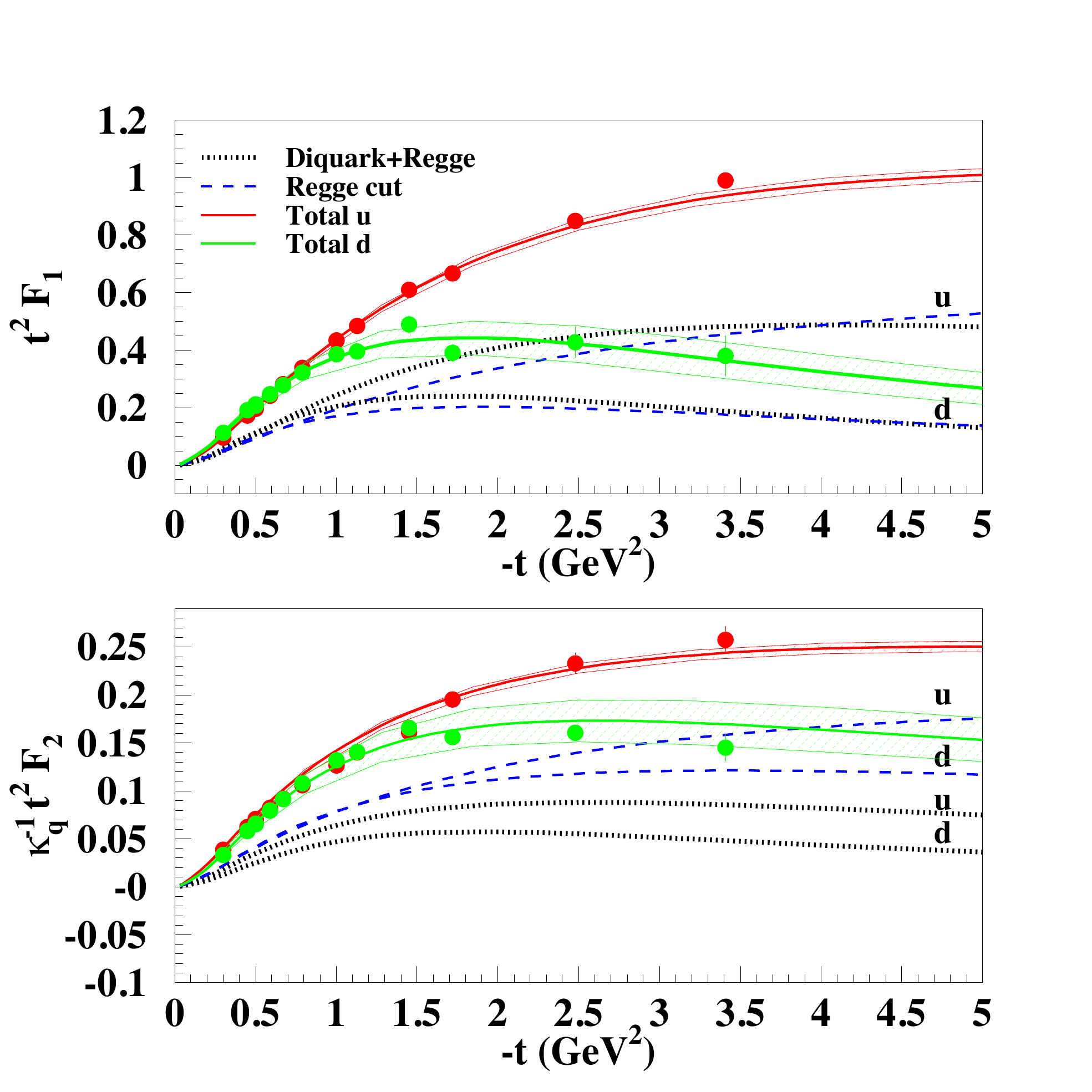}
\caption{(Color online) $t^2 F_1^q$, $q=u,d$ (above) and $\kappa_q^{-1} t^2 F_2^q$ (below) plotted vs. $-t$, as obtained from our parametrization using  Eq.(\ref{fit_form}). 
Experimental data from Ref.\cite{Cates}. The dotted lines correspond to the diquark contribution, Eq.(\ref{RG1}); the dashed lines to the Regge contribution, Eq.(\ref{RG2}). }
\label{fig11}
\end{figure}

\begin{figure} 
\includegraphics[width=9.5cm]{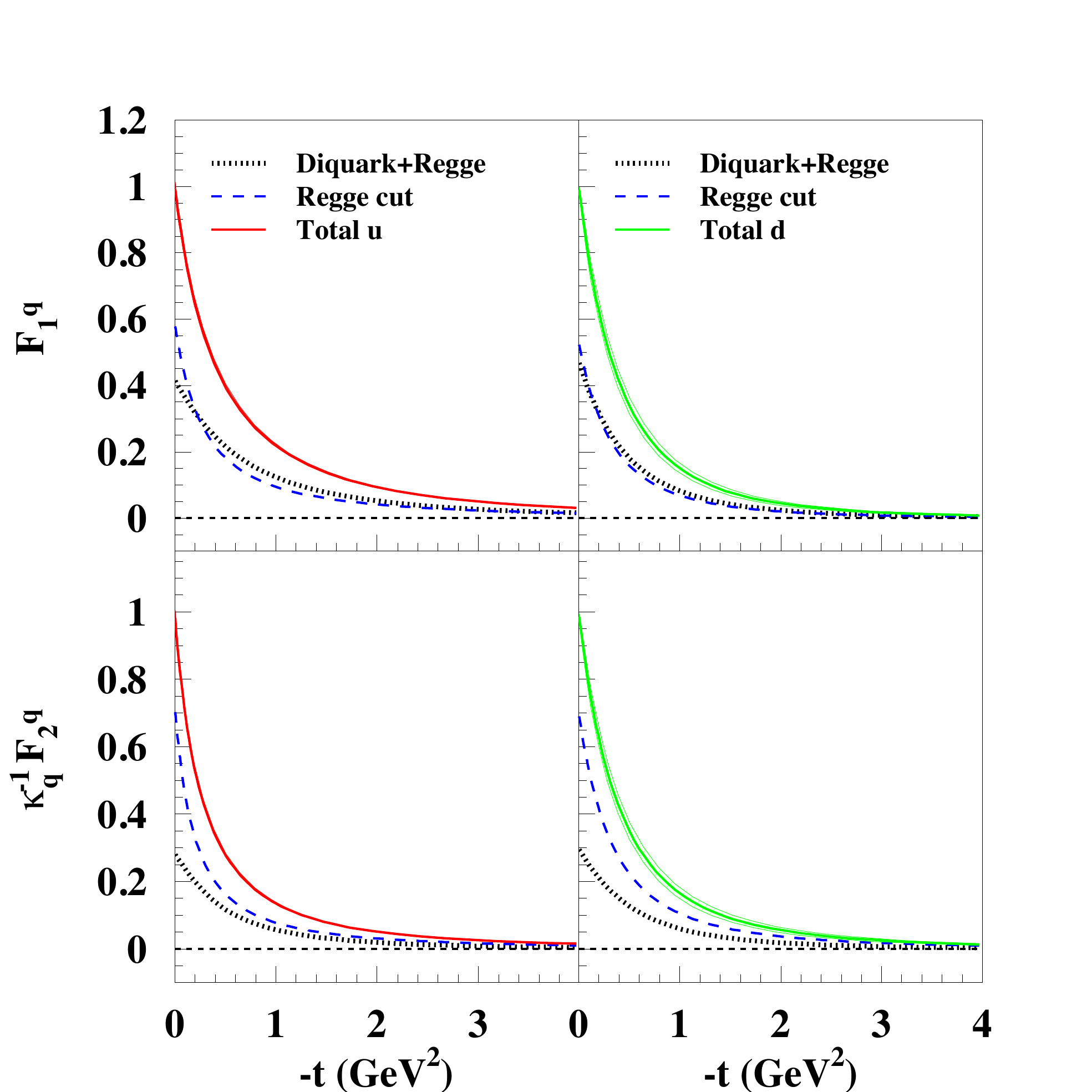}
\caption{(Color online)  Form factors, $F_1^q$, (above), and $\kappa_q^{-1} F_2^q$ (below). $u$ and $d$ quarks contributions displayed in the left and right panels, respectively. 
 The dotted lines correspond to the diquark contribution, Eq.(\ref{RG1}); the dashed lines to the Regge contribution, Eq.(\ref{RG2}). }
\label{fig12}
\end{figure}

This can also be seen in Fig.\ref{fig11} and Fig.\ref{fig12} where the single quark, and rescattering contributions to the form factors are presented along with the total contribution. In Fig.\ref{fig12} on has a better view of the small $t$ behavior.  

\subsubsection{Summary}
To summarize, 
we notice first of all that in our diquark model the $t$-dependence of both GPDs and the form factors
originates from two contributions describing scattering from a single quark (including a single Regge exchange dominating low $x$ and $t\approx 0$, Figs.\ref{fig1} and \ref{fig8}a),
and diquark correlations (Fig.\ref{fig8}b) which we effectively take into account through Regge re-interactions, or Regge cuts. The latter appear only through the reggeization procedure. The single quark scattering terms involve, therefore, both a fixed mass diquark contribution, Eqs.({\ref{G}) and the $t$-independent Regge term, $\approx x^{-\alpha_q}$, in Eq.(\ref{regge}) .
  
An important outcome of our analysis is that the re-interaction components are necessary for a quantitative description of the form factors 
in a wide range of $t$.  
At low $t$ single quark scattering dominates $F_1^u$  as expected (Fig.\ref{fig12}),  
while for $F_1^d$ the contributions of the Regge re-interactions and single quark terms are comparable. 

Scattering from a single, non-interacting $u$ or $d$ quark leaves either a $ud$ diquark with spin $S=0,1$, or a $uu$ diquark system with spin $S=1$ as spectators. 
The angular momentum structure of the $J^P=0^+$ (scalar) and $J^P=1^+$ (axial vector) configurations allows us to obtain distinct predictions for the $u$ and $d$ quarks contributions using an SU(4) symmetric wave function for the proton,
\begin{eqnarray}
\label{Fu}
F^u(X,\zeta,t) & = &  \frac{3}{2} F^{S=0}(X,\zeta,t) + \frac{1}{2} F^{S=1}(X,\zeta,t)    \\
\label{Fd}
F^d(X,\zeta,t) & = & F^{S=1}(X,\zeta,t).
\end{eqnarray} 
After performing the azimuthal angle integration in Eqs.(\ref{H},\ref{E}) one can see that two $\Delta_T$ dependent terms survive in the numerator for $H$, while $E$ depends on $\Delta_T$ only in the denominator. 
This yields a steeper $t$ dependence in the Pauli form factor. 
The flavor dependence of both the  Dirac and Pauli form factors is governed by the precise values of the masses entering the term \[{\cal M}_q^2(x)  =  x M^2 - \frac{x}{1-x} M_{X}^{q \, 2} -M_{\Lambda}^{q \, 2}, \] 
in Eqs.(\ref{H}{,\ref{E}). 

The masses values obtained in different diquark models \cite{hybrid_even,Roberts,BCR}, are reported in Table \ref{table:mass} and Table \ref{table:diquarkcutoff}. In Table \ref{table:mass} we show the quark and diquark masses. In Table  \ref{table:diquarkcutoff} we show the value of the diquark form factor mass term, $M_\Lambda^q$, Eq.(\ref{coupling}), along with the corresponding radii calculated as \cite{Roberts},
\begin{equation}
r_{diq}^q = \frac{\sqrt{6}}{M_\Lambda^q}
\end{equation}
The diquark form factor used in our model, Eq.(\ref{coupling}) plotted vs. $k\equiv k_\perp$, for the different flavor components is shown in Fig.\ref{fig13}.  We notice only a slight flavor dependence of these terms. This trend can be seen also by comparing the value of the radii  In Table  \ref{table:diquarkcutoff}.

From Table \ref{table:mass} one can see that for all models we find  that $M_X^u \lesssim M_X^d$.  This produces in turn, a slightly steeper fall with $t$  for the $d$ quarks than for the $u$ quarks, occurring in a similar way for both $F_1^q$ and $F_2^q$. 
%
\begin{figure}
\includegraphics[width=8.5cm]{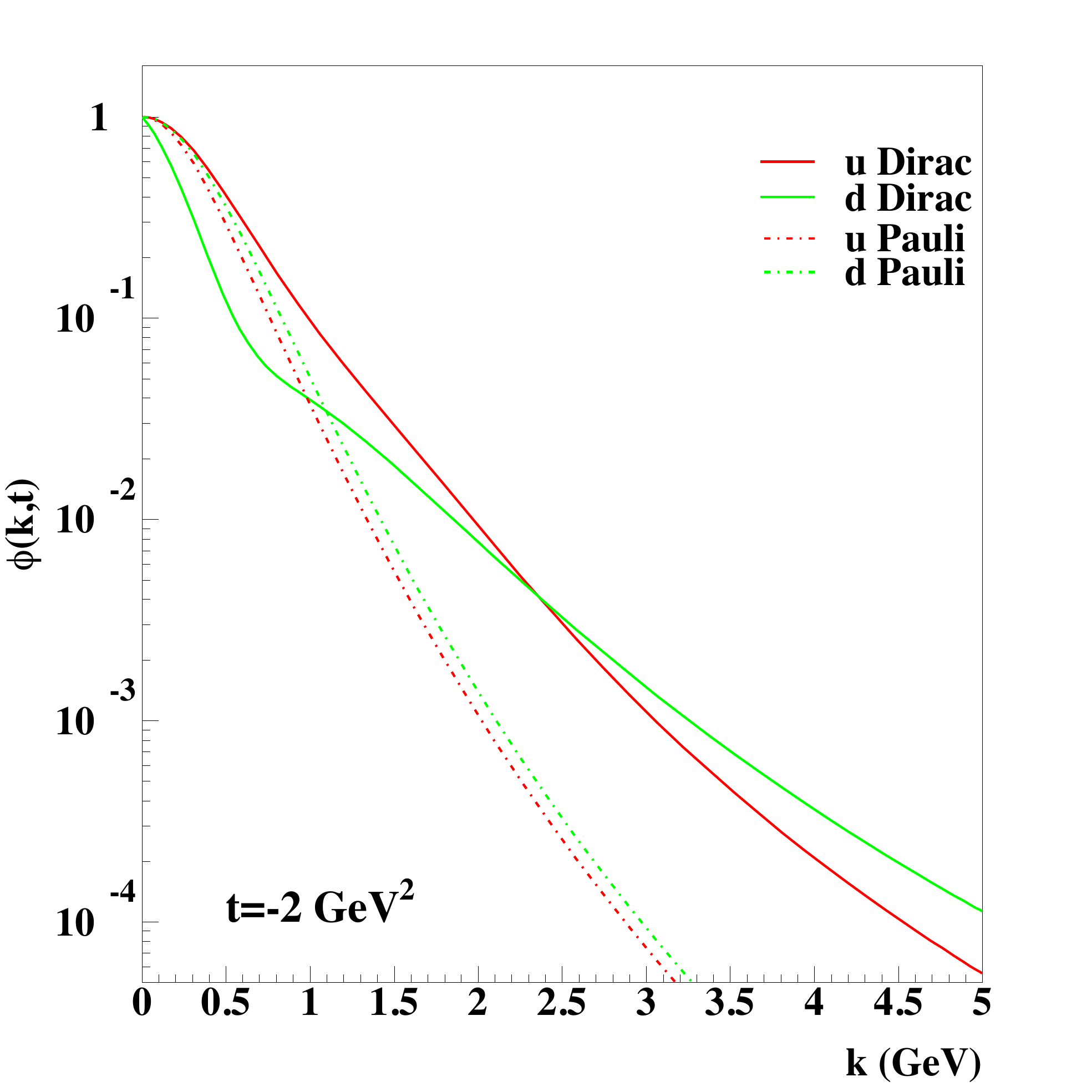}
\caption{(Color online) Diquark vertex function contributions to the Dirac and Pauli form factors obtained by integrating Eq.(\ref{coupling}) over $x$ at fixed $t=-2$ GeV$^2$, plotted vs. $k\equiv \mid {\bf k}_\perp\mid$.}
\label{fig13}
\end{figure}

The behavior of the fixed mass diquark term  can be seen in Fig.\ref{fig11} where the single quark scattering contribution was plotted separately from the rescattering term.
In this respect, our findings  are in agreement with the trend for the flavor dependence of the form factors predicted in Ref.\cite{Roberts}.   
\begin{table}
\begin{tabular}{|c|c|c|c|c|c|}
\hline
\hline
Reference : mass (GeV) & $ m_u $& $M_{X=ud}^u (J=0^+) $ &$ m_d $  & $M_{X=uu}^d (J=1^+)$ & $M_{X=ud}^u (J=1^+)$  \\
\hline
\hline  
GGL \cite{hybrid_even} & $0.420$ & $0.604$ & $0.275$ & $0.913$ & $0.604$  \\
\hline 
Cloet et al. \cite{Roberts} &   $0.33$ &   $0.7-0.8 $  &  $0.33$   &    $0.9-1.0$ & $0.9 -1.0$  \\ 
\hline 
BCR \cite{BCR} & $0.3 $   & $0.822$   & $0.3$ &   $0.890$ & $1.492$ \\
\hline
\end{tabular} 
\caption{%
\label{table:mass} Values of the quark and diquark masses in GeV calculated in several models. The quark masses values in \cite{Roberts}  correspond to the constituent quarks limit in this model.}
\end{table}
However, in order  to quantitatively explain  the difference in the $u$ and $d$ quarks' behavior  we emphasize that one needs to go beyond single quark scattering and consider diquark correlations/Regge re-interactions. The latter allow the Regge term to be present at larger $t$. In fact, as shown in Fig.\ref{fig7},  the diquark correlations/Regge cuts extend, in fact,  the validity of the Regge model to this region \cite{ColKea,PDBCol}. At large $t$, in the multi-GeV region, the form factors are dominated by the large $x$ components of the GPDs, Fig.\ref{fig10} (see also Refs.\cite{Rad98,Radyushkin_par}).

Notice also that in our model the $d$ quark form factor is not predicted to become negative when extrapolated to larger momentum, while a flattening of its slope in $-t$ 
occurs.

\subsection{Scale Dependence}
GPDs are dynamical quantities that  depend on the value of the scale, $Q^2$,  of the deep inelastic process that is used to measure contrarily to their 
first moments which are given by the form factors. The form factors connect to GPDs at any value of the scale. It is therefore important to determine whether the flavor dependence interpreted so far in terms of their partonic substructure at the initial scale, $Q_o^2$, changes with the scale.   
This can be evaluated by using the Perturbative QCD (PQCD) evolution equations for the GPDs at a given order \cite{MusRad,GolMar}. In our model we use the expressions for the kernels at Leading Order (LO)  with $\Lambda_{QCD}^{N_f=4}= 0.215$, and $Q_o^2 \approx 0.1$ GeV$^2$. 
Results of PQCD evolution are given in Fig.\ref{fig:scale} and Fig.\ref{fig:scale_2}. In  Fig.\ref{fig:scale_2}, in particular, we show the ratio,
\begin{eqnarray}
R_q = \frac{F_1^q(t,x_{MAX};Q^2)}{F_1^q(t)},
\end{eqnarray}
where,
\begin{eqnarray}
F_1^q(t,x_{MAX}; Q^2) = \int_{-1}^{x_{MAX} } dx \, H_q(x,0,t;Q^2),
\end{eqnarray}
$q=u,d$.
From Fig.\ref{fig:scale} one can see that although accounting for the $Q^2$ dependence of the GPDs changes the shape of the curves in a predictable way, {\it i.e.} moving ``strength" to lower values of $x$, this does not affect the flavor dependence interpretation of the different components of our model which keeps on being valid at larger scales. In other words, the two peaks describing the single quark scattering and interactions persists, and they are located at different $x$ values for the $u$ and $d$ quarks, respectively.
One can see this also from Fig.\ref{fig:scale_2} where the $d$ quarks ratio saturates faster then the $u$ quarks', the latter being dominated by higher $x$ components (Fig.\ref{fig10}).  However, Fig.\ref{fig:scale_2} also shows that both the $u$ and $d$ quarks form factor components are governed by increasingly lower $x$ components as the scale of the process increases. The dynamical properties of GPDs should be taken into account in order to connect them to form factors. 

\begin{table}
\begin{tabular}{|c|c|c|c|c|c|c|}
\hline
\hline
Reference  & $M_\Lambda^{ud} (J=0^+) $  & $r^{ud} (fm)$ & $M_\Lambda^{uu} (J=1^+)$ & $r^{uu}$ (fm) & $M_\Lambda^{ud} (J=1^+)$  & $r^{ud}$ (fm) \\
\hline
\hline  
GGL \cite{hybrid_even} & $1.018$ &  0.330 & $0.860$ & 0.390  & $0.860$  & 0.390 \\
\hline 
Cloet et al. \cite{Roberts} &   $0.479$ &   $0.7 $  & -   &  - & 0.419    &    $0.8$ \\ 
\hline 
BCR \cite{BCR} & $0.609$ &  0.551 & $0.376$ & 0.892  & $0.716$ & 0.469  \\
\hline
\end{tabular} 
\caption{%
\label{table:diquarkcutoff} Values of the diquark masses cutoffs in GeV and corresponding diquark configurations radii in fm, as calculated in several models.}
\end{table}
%
\begin{figure}
\includegraphics[width=9.0cm]{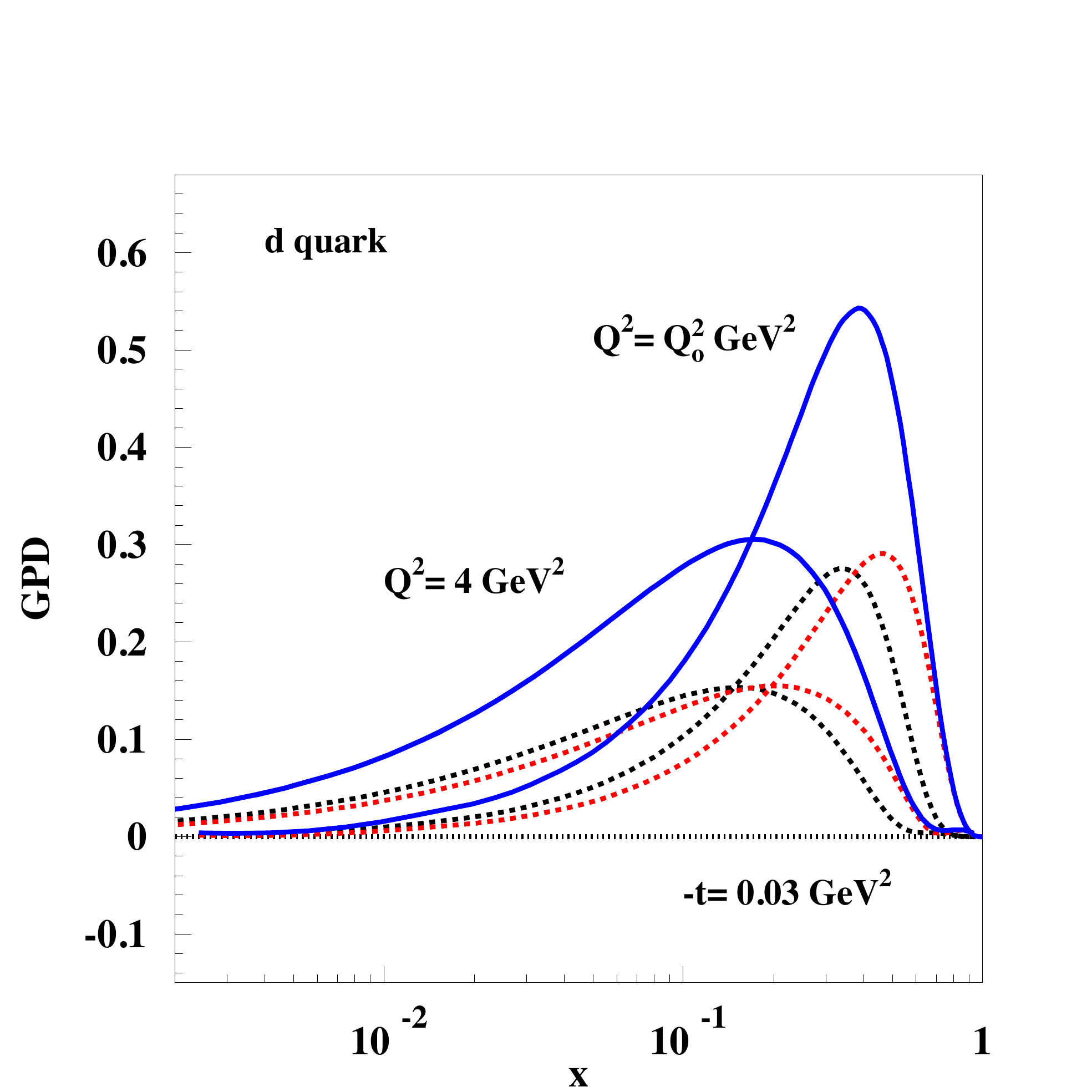}
\caption{(Color online) GPD $x H_d(x,0,t; Q^2)$ plotted vs. $x$, for $t=-0.03$ GeV$^2$, and two values of the process' scale: $Q^2= 4$ GeV$^2$, and $Q_o^2=0.1$ GeV$^2$. The diquark + Regge and the Regge re-interactions terms are shown separately, following the notation of Fig.\ref{fig9}. Analogous results are obtained for $F_1^u$, and for the Pauli form factors.}
\label{fig:scale}
\end{figure}
\begin{figure}
\includegraphics[width=8.0cm]{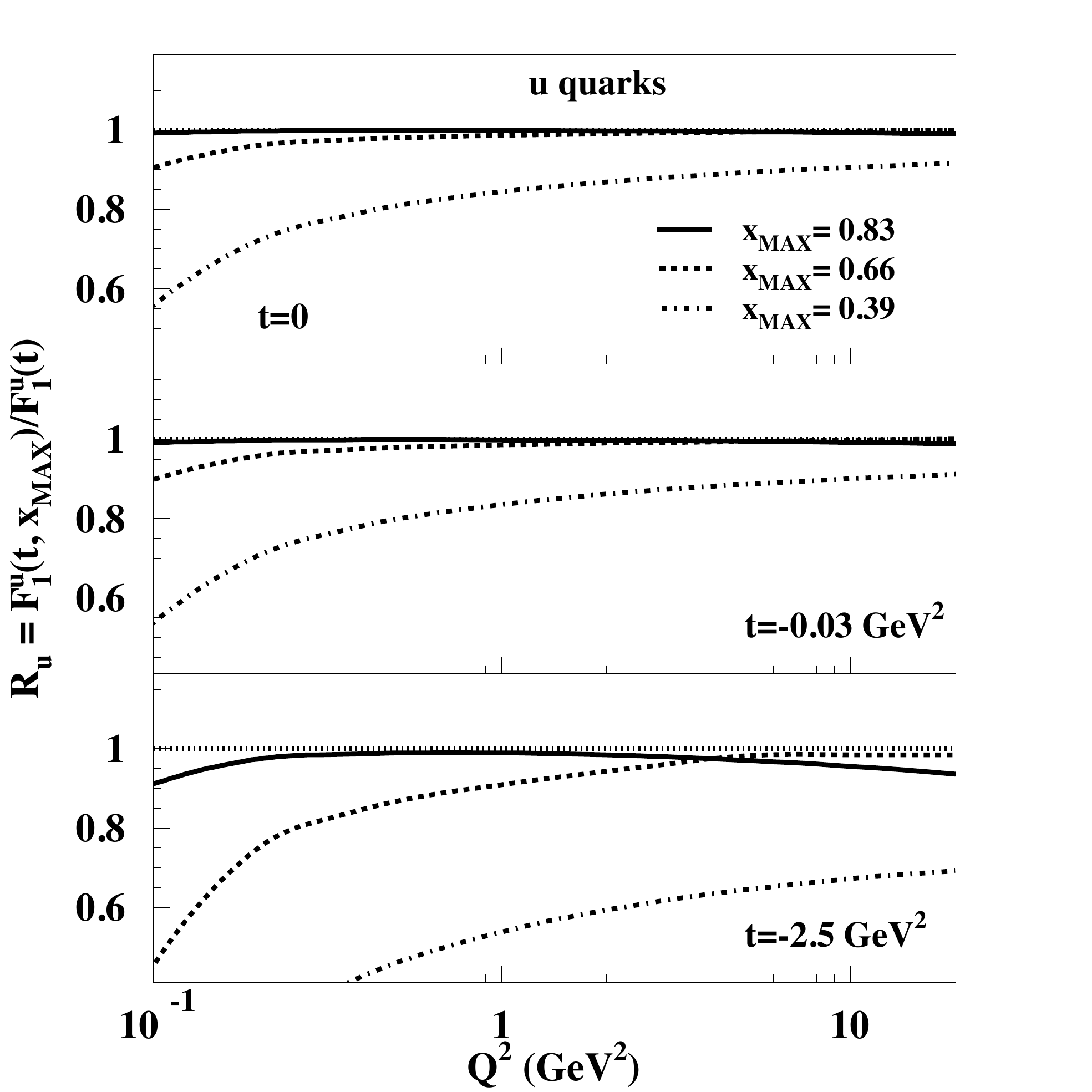}
\includegraphics[width=8.0cm]{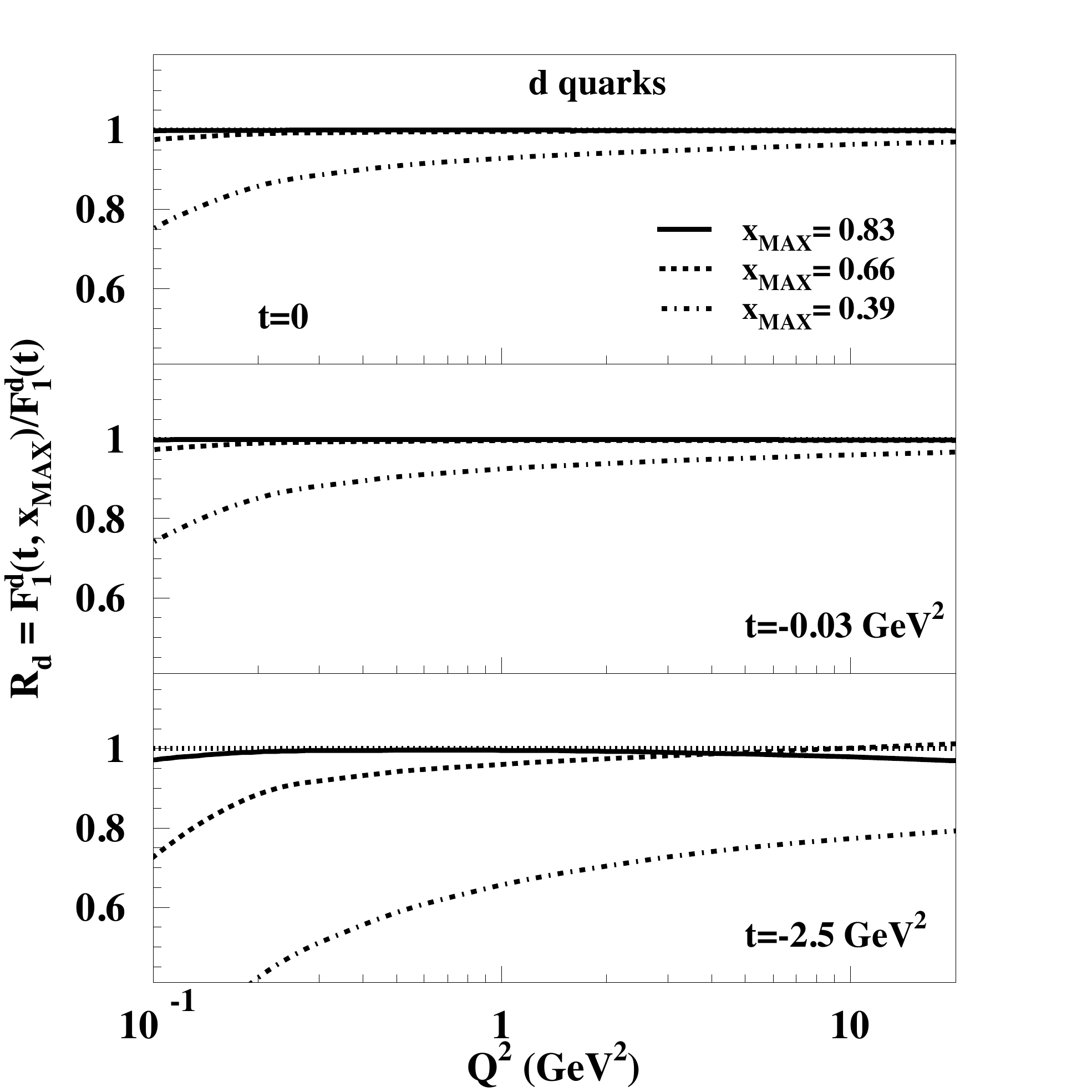}
\caption{GPD $x H_d(x,0,t; Q^2)$ plotted vs. $x$, for $t=-0.03$ GeV$^2$, and two values of the process' scale: $Q^2= 4$ GeV$^2$, and $Q_o^2=0.1$ GeV$^2$. The diquark + Regge and the Regge re-interactions terms are shown separately, following the notation of Fig.\protect\ref{fig9}.}
\label{fig:scale_2}
\end{figure}

\subsection{Transverse Space Analysis}
All of the questions discussed in the previous Sections impact the transverse radial dependence that can be deduced from GPD based analyses.The connection between form factors and Fourier transforms of GPDs was studied quantitatively in Ref.\cite{LiuTan04,arrmil,Miller_1}. In particular, in Ref.\cite{arrmil} a partonic interpretation was given of the negative central charge density of the neutron in terms of  the correlation between the dominance of $d$ quarks at large $x$ and their transverse radii.   

In what follows we study the flavor dependence of the transverse densities obtained from the form factors, Eqs.(\ref{FT}) using our model. 
More detailed studies addressing also the $Q^2$ dependence in transverse coordinates space will be considered in  \cite{Osvaldo_prep}.

The Fourier transforms of GPDs with respect to ${\bf \Delta}_\perp$ define the parton density distributions at a transverse position
{\bf b} for a given longitudinal momentum fraction, $x$, namely  two dimensional
distributions in the transverse plane with respect to the proton's direction of motion. 
We can therefore connect each form factor component to hadronic distances from the proton's center of momentum defined as \cite{arrmil},
\begin{eqnarray}
\label{radii}
\langle b^2 \rangle^{q}_1  & = & \int_0^1 dx \int \frac{d^2 b}{(2\pi)^2} \, \rho^q_1(x,b) \, b^2  \nonumber \\
& = &  \int_0^1 dx \int \frac{d^2 b}{(2\pi)^2} \left[ \int  d^2 \Delta_\perp \,\, [H^q_{diq}(x,0,\Delta_\perp^2)+H^q_R(x,0,\Delta_\perp^2)] e^{i {\bf b} \cdot \Delta_\perp}  \right] b^2  \nonumber \\
& = & \langle  b^2 \rangle ^q_{1 \, diq}  +  \langle  b^2 \rangle ^q_{1 \, R},  \\
\langle b^2 \rangle^q_2   & =  & \int_0^1 dx \int \frac{d^2 b}{(2\pi)^2} \, \rho^q_2(x,b) \, b^2  \nonumber \\
& = & \int_0^1 dx \int \frac{d^2 b}{(2\pi)^2}  \left[  \int  d^2 \Delta_\perp \,\,  [E^q_{diq}(x,0,\Delta_\perp^2)+E^q_R(x,0,\Delta_\perp^2)] e^{i {\bf b} \cdot \Delta_\perp} \right] b^2 \nonumber \\
& = & \langle  b^2 \rangle^q _{2 \, diq}  +  \langle  b^2 \rangle ^q_{2 \, R} .
\end{eqnarray}
In Fig.\ref{fig16} we show the density integrated over $x$, 
\begin{equation}
\rho^q_{1(2)}(b) = \int_0^1 dx  \, \rho^q_{1(2)}(x,b)
\end{equation}
 for all components. 
In Table \ref{table:radii} we show the values of the radii, $b_{1(2) \, diq(R)}^q = [\langle b^2 \rangle_{1(2) \, diq(R)}^q]^{1/2}$  for the single mass diquark ($diq$) and reggeized ($R$) components in our model. 

 From Table \ref{table:radii}  and Fig.\ref{fig16}  one can see that in average, the $u$ quark transverse distance is smaller than the $d$ quark's, thus confirming a picture similar to the one proposed in \cite{Roberts}  although using different symmetry properties since the radii there include also the longitudinal spatial component. Based on our previous discussion, one can easily relate the values in Table \ref{table:radii} 
to the transverse momentum distribution, regulated by the vertex function in Eq.(\ref{coupling}),  which is displayed in Fig.\ref{fig13}. We conclude that although a space coordinates description gives us very useful information \cite{Miller_1,CloetMiller}, it is probably too simplistic to give an interpretation of the flavor dependence in terms of {\it average} quark distances inside the proton, essentially due to re-interactions (as, in fact, also noticed in \cite{Roberts}). The effect of re-interactions is, however, very interesting to explore in itself since it allows us to estimate the distance of  a non point-like two-quark hadronic component containing either the struck $u$ or $d$ quark  from the proton's center of momentum. Our analysis points at interesting differences in the flavor dependence behavior of the Dirac vs. Pauli form factors. The behavior of the form factors at larger $t$ is reflected in the behavior of $\rho$ as $b \rightarrow 0$. In particular, we notice that the data on $F_2$ at intermediate $t$ show a smaller relative deviation of the $d$ quarks from the $u$ quarks, or a harder $d$ quark component (see also Fig.\ref{fig10}).   Future data at larger momentum transfer will help validating this interpretation. A more detailed study that emphasizes the description in coordinate space is on its way \cite{Osvaldo_prep}.

\begin{figure}
\includegraphics[width=8.0cm]{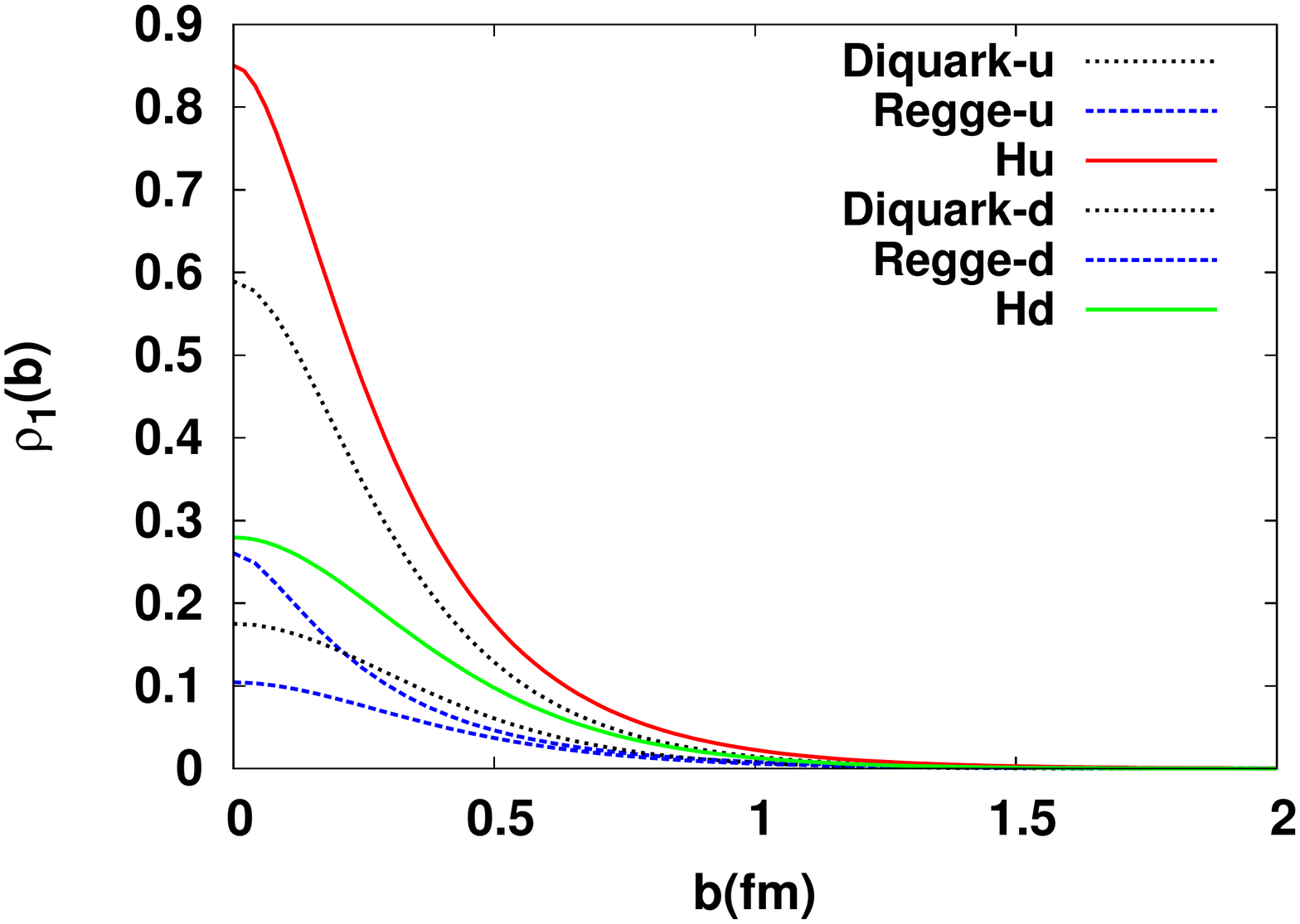}
\includegraphics[width=8.0cm]{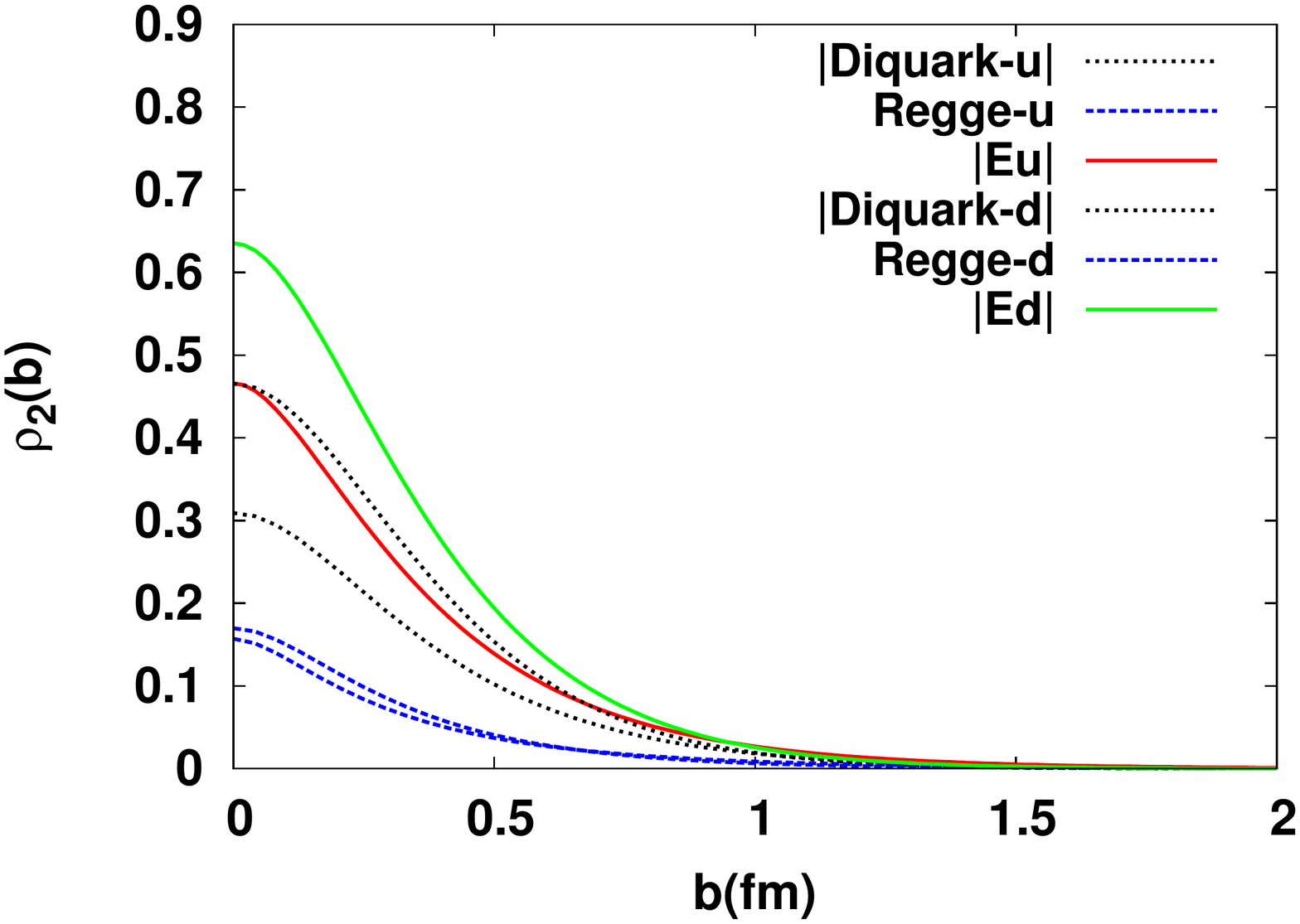}
\caption{(Color online) Left Panel: Integrated over $x$ parton density distribution in transverse space  for the Dirac form factor, displaying  all components of our model. Right Panel: same as left, for the Pauli form factor.}
\label{fig16}
\end{figure}
\begin{table}
\begin{tabular}{|c|c|c|c|c|c|c|}
\hline
\hline

Flavor    &  $b^q_{1 \, {diq}}$    & $b^q_{1 \, R}$ & $b^q_1$   &  $b^q_{2 \, diq}$    & $b^q_{2 \, R}$  & $b_2^q$      \\ 
\hline
\hline
$u$	&0.475	&0.385	&0.612	&0.527	&0.447	&0.691 \\	
\hline
$d$	&0.450	&0.425	&0.619	&0.745	&0.463	&0.877 \\
\hline
\hline
\end{tabular}
%
\caption{%
\label{table:radii} 
Radii in fm per quark flavor, GPD type, and model components as described in the text. The notation is: 
$b^q_{1(2) \, diq} \equiv  [\langle  b^2 \rangle ^q_{1(2) \, diq}]^{1/2} $, $b^q_{1(2) \, R} \equiv  [\langle  b^2 \rangle ^q_{1(2) \, R}]^{1/2} $, $b^q_{1(2)} \equiv [ \langle  b^2 \rangle ^q_{1(2)}]^{1/2} $. }
\end{table} 

\subsection{Flavor Dependent Angular Momentum}
Finally, using the parameters from our analysis of the flavor dependence of the proton form factors we can make a prediction for the values of the quarks total angular momentum, $J_q$, $q=u,d$. $J_q$ is defined in Ji's sum rule as \cite{GPD2},
\begin{equation}
J_q = \frac{1}{2} \int_{-1}^{1}  dx \, x (H_q(x,0,0; Q^2) + E_q(x,0,0; Q^2) ),
\end{equation}
and the quarks orbital angular momentum obtained as,
\begin{equation}
L_q =  J_q - \frac{1}{2} \int_{-1}^{1}  dx \,  \tilde{H}_q(x,0,0; Q^2)
\end{equation}
where $Q^2$ is the process' scale.
Our results obtained evolving all GPDs at leading order \cite{hybrid_even} are shown in Figs. \ref{fig17},\ref{fig18} and in Table \ref{table:Jq}.
\begin{figure}
\includegraphics[width=9.0cm]{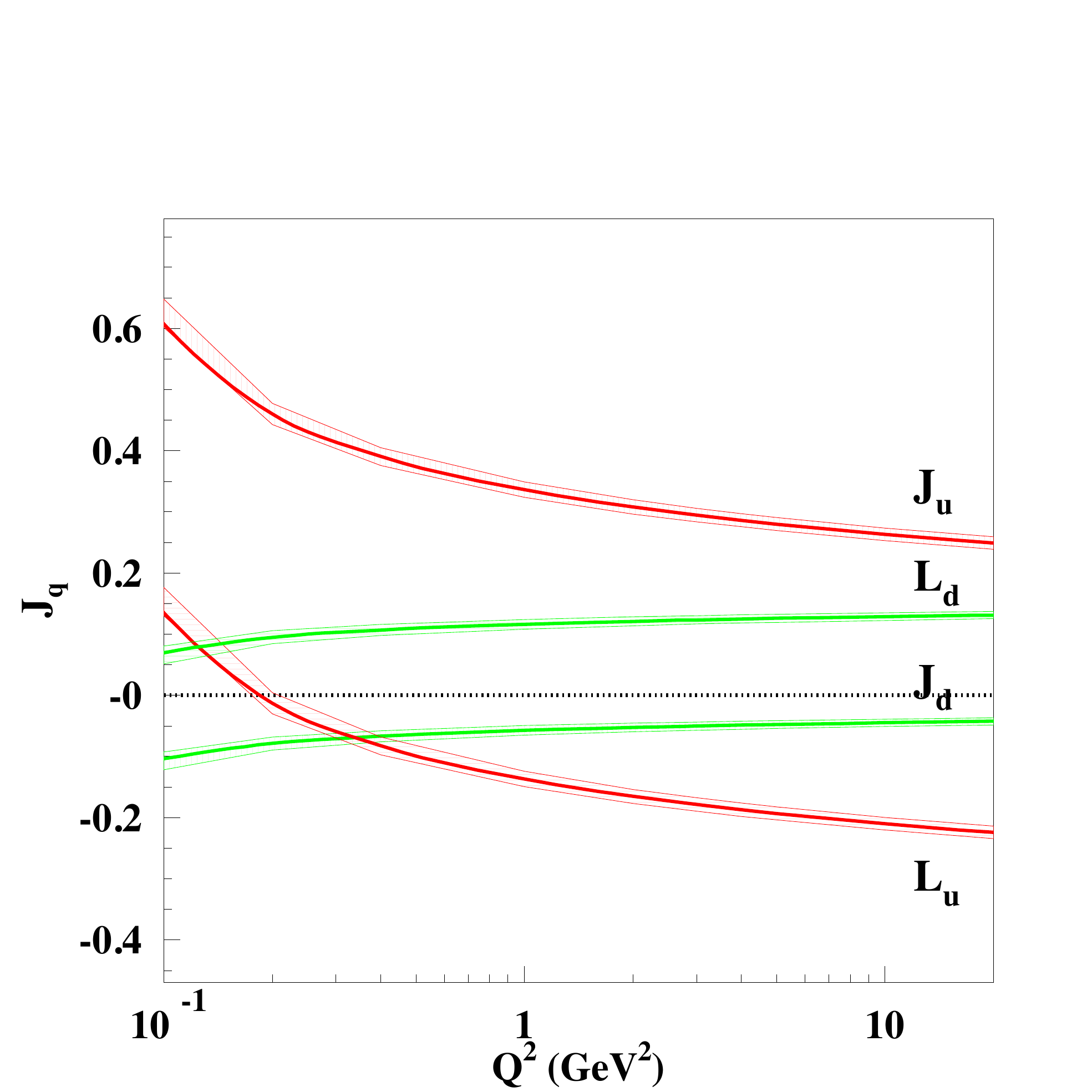}
\caption{(Color online) Quarks angular momentum, $J_q$, and orbital angular momentum, $L_q$, plotted vs. the scale $Q^2$.}
\label{fig17}
\end{figure}
\begin{figure}
\includegraphics[width=9.0cm]{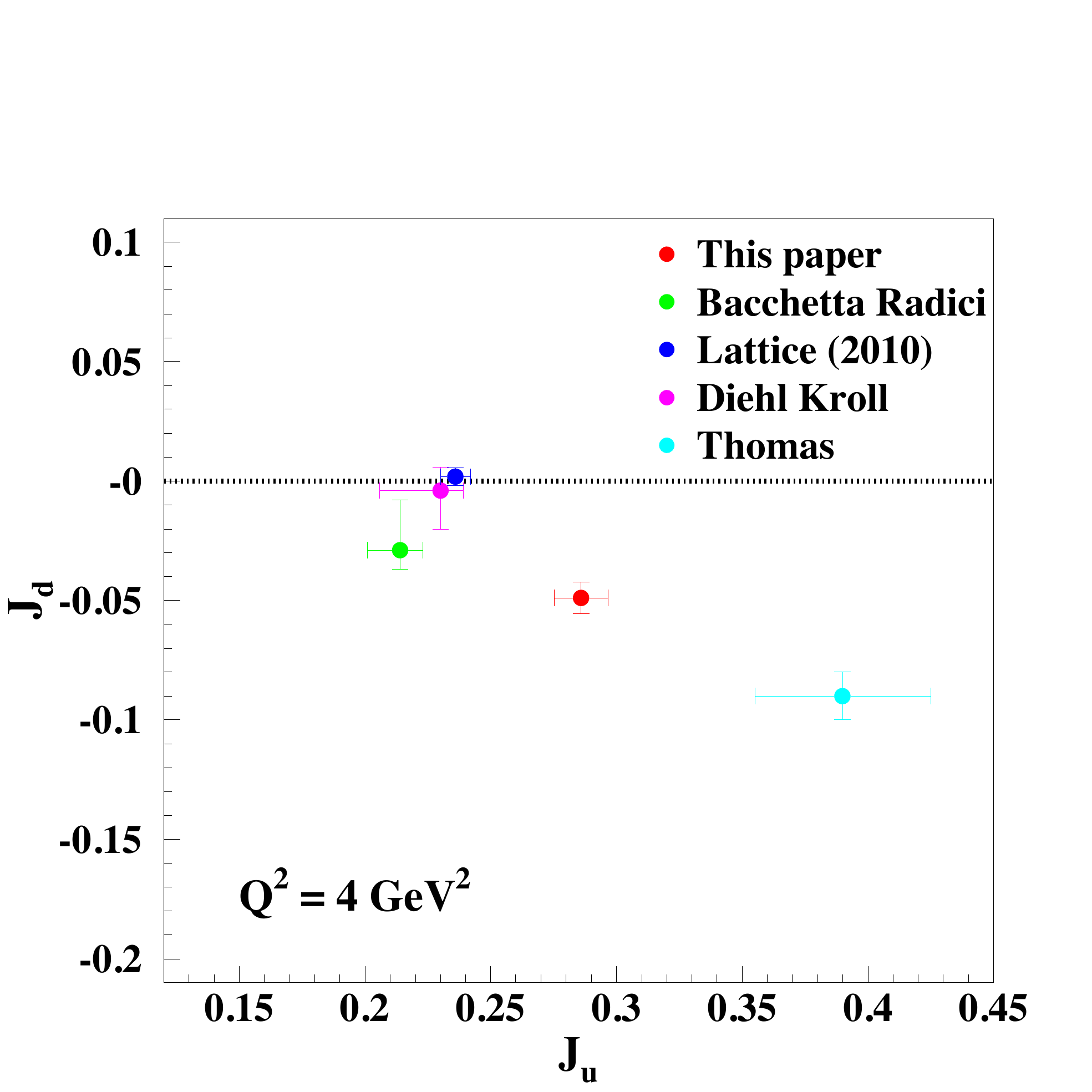}
\caption{(Color online) $J_u$ vs. $J_d$ obtained from our parametrization which is constrained by the flavor separated Dirac and Pauli form factors, compared to other determinations including  theoretical uncertainty:
a similar analysis in Ref.\cite{Diehl_Kroll13}, the model calculation in Ref.\cite{Thomas}, the model dependent analysis including Transverse Momentum Distributions (TMDs) data of Ref.\cite{BR}, and the most recent lattice QCD evaluation \cite{Lattice,Hagler:2009}.} 
\label{fig18}
\end{figure}
\begin{table}
\begin{tabular}{|c|c|c|c|c|c|}
\hline
\hline
Reference   &  This paper   & LHPC \cite{Lattice} & Thomas \cite{Thomas}  &  TMDs \cite{BR}    & Diehl \& Kroll \cite{Diehl_Kroll13}        \\ 
\hline
\hline
$u$  &  0.286 $\pm$  0.107 &    0.236 $\pm$ 0.0018&   0.390 $\pm$ 0.035 &   0.214  $\begin{array}{c} +0.009  \\ -0.013 \end{array}$    &   0.230  $\begin{array}{c} + 0.009 \\ -0.024 \end{array}$ \\
\hline
$d$  &  -0.049  $\pm$ 0.007 &   0.006 $\pm$ 0.0037 &  -0.09  $\pm$ 0.01&  -0.029  $\begin{array}{c} + 0.021 \\ -0.008 \end{array}$     &    -0.004  $\begin{array}{c} + 0.010 \\ -0.016 \end{array}$ \\
\hline
\hline
\end{tabular}
\caption{\label{table:Jq} %
Values of angular momentum, $J_u$ and $J_d$, at $Q^2= 4$ GeV$^2$, obtained in various approaches: our parametrization which is constrained by the flavor separated Dirac and Pauli form factors, compared to other determinations including  theoretical uncertainty:
from a similar analysis in Ref.\protect\cite{Diehl_Kroll13}, from a model calculation \protect\cite{Thomas}, from a model dependent analysis including Transverse Momentum Distributions (TMDs) data \protect\cite{BR}, and from the most recent lattice QCD evaluation \protect\cite{Lattice,Hagler:2009}. }
\end{table} 
The model dependence in various calculations arises entirely from the GPD $E$, since the second moment of $H$ is precisely constrained by deep inelastic scattering measurements. It is interesting to notice a discrepancy with the analysis of Ref.\cite{Diehl_Kroll13} where similar constraints from the nucleon form factors were used. We conclude that even if the new precise measurements of the form factors reduce the uncertainty in the GPDs, in order to obtain angular momentum, direct measurement of $E$ through DVCS type experiments are mandatory.

\section{Conclusions}
\label{sec:4}
In conclusion, we used a GPD based approach as a way to understand the behavior of the $u$ and $d$ quarks components of the nucleon form factors. The GPDs were evaluated using a reggeized quark-diquark model whose parameters are fixed to simultaneously fit the deep inelastic limit, the nucleon form factors, and DVCS data. 
%
A unified picture of the Regge and diquark contributions can be given using duality arguments.  The Regge terms include a component that corresponds to diquark correlations in the nucleon.  

Reggeization, through a spectral distribution, $\rho(M_X^2)$, for the diquark system's mass,  accounts for the more complex correlations that appear at large mass values of the diquark system. Reggeization is the source of diquark correlations in our model. At low mass values the diquark system behaves as two quarks with spin $J=0^+,1^+$, and scattering occurs from a single quark within the impulse approximation.

A first outcome is that  the new highly precise form factor data produce much improved constraints on our GPDs parameters.
The interpretation of the flavor dependence of the data lies in the non-perturbative structure of both the Regge and quark-diquark terms. It is a subtle combination of effects that cannot be ascribed to a single, simply motivated mechanism. 

In a quark-diquark scattering picture flavor dependence arises from the difference in masses between the axial vector (dominating the $d$ quarks) and the scalar (dominating the $u$ quark) diquark components, which in turn define different size average radii for the two flavors. 
However, we found out that diquark re-scattering mechanisms are important, being responsible for a further shift to large $x$ values which occurs in different proportions for the $u$ and the $d$ quarks. This in turn can be explained in terms of the types of $t$-channel quantum numbers (or  reggeons, according to a duality picture) being exchanged, rather than directly in terms of mass values.
In connecting to GPDs it is important to take into account the scale dependence of the process. We found that although through PQCD evolution the form factors expectedly  relate to larger $x$ components of the GPDs at low $Q^2$ than at larger $Q^2$, this trend occurs similarly for the $u$ and $d$ quarks, and it is therefore not flavor dependent.  
  
Through the concept of GPDs the Regge and diquark mechanisms are realized  correspondingly in coordinate space, in the transverse plane where our description reflects the behavior of the form factors.  In the GPD picture one can also study the internal spatial distribution and size of these components.

Finally, by fixing the GPD parameters using the flavor separated form factor data we could improve on our estimate of the values of their second moment, which measures the proton's angular momentum. 
 
The newly available flavor separated form factor data at large $t$ stimulated this work. Further studies exploring a possibly important role of diquark/few-parton correlations inside the proton will be carried out in the near future. 

\vspace{0.5cm}
We thank Gordon Cates for helpful discussions, and Oscar Rondon for comments on the treatment of experimental errors in form factors analyses. This work has been supported by the U.S. Department
of Energy grants DE-FG02-01ER4120 (J.O.G.H., , K.K., S.L.), and DE-FG02-92ER40702  (G.R.G.).


\end{document}